\documentclass[sigconf]{acmart}

\usepackage{booktabs} 
\usepackage{flushend}

\setcopyright{acmcopyright}

\acmDOI{10.475/123_4}

\acmISBN{123-4567-24-567/08/06}

\acmConference[TheWebConf 2019]{May 13-17}{San-Francisco}{USA}
\acmYear{2019}
\copyrightyear{2018}

\begin{document}
\title{Anomaly detection in the dynamics of web and social networks}
\renewcommand{\shorttitle}{Anomaly detection using unsupervised learning}

\author{Volodymyr Miz, Benjamin Ricaud, Kirell Benzi, and Pierre Vandergheynst}
\affiliation{%
  \institution{Ecole Polytechnique F\'ed\'erale de Lausanne}
  \city{Lausanne}
  \state{Switzerland}
}
\email{first.last@epfl.ch}

\renewcommand{\shortauthors}{V. Miz et al.}

\begin{abstract}

In this work, we propose a new, fast and scalable method for anomaly detection in large time-evolving graphs. It may be a static graph with dynamic node attributes (e.g. time-series), or a graph evolving in time, such as a temporal network. We define an anomaly as a localized increase in temporal activity in a cluster of nodes. The algorithm is unsupervised. It is able to detect and track anomalous activity in a dynamic network despite the noise from multiple interfering sources. 

We use the Hopfield network model of memory to combine the graph and time information. We show that anomalies can be spotted with a good precision using a memory network. The presented approach is scalable and we provide a distributed implementation of the algorithm.

To demonstrate its efficiency, we apply it to two datasets: Enron Email dataset and Wikipedia page views. We show that the anomalous spikes are triggered by the real-world events that impact the network dynamics. Besides, the structure of the clusters and the analysis of the time evolution associated with the detected events reveals interesting facts on how humans interact, exchange and search for information, opening the door to new quantitative studies on collective and social behavior on large and dynamic datasets.

\end{abstract}

%
%



\maketitle

\section{Introduction}\label{Introduction}

In recent years, the Web has significantly affected the way people learn, interact in social groups, store and share information. Apart from being an essential part of modern life, social networks, online services, and knowledge bases generate a massive amount of logs containing traces of global online activity on the Web. 
Most of this data is related to the standard activity of the users. However, the larger these logs become, the harder it is to detect deviations from normal behavior in the network. Localization of these anomalies becomes even more difficult because of the continuous expansion and dynamic nature of these networks.
Hence, in recent years, anomaly detection has become an important field of research focusing on this problem~\cite{chandola2009anomaly}, \cite{ranshous2015anomaly}.



In this paper, we present a scalable anomaly detection approach for spatio-temporal datasets such as user activity logs of web and social networks. The approach focuses on anomaly detection in the collective behavior of users in such networks. Our approach introduces the following novelties that distinguish it from the state-of-the-art.
 

1) The method bridges the gap between graph anomaly detection~\cite{ranshous2015anomaly,akoglu2015graph,aggarwal2014evolutionary} and spatio-temporal data mining~\cite{atluri2018spatio}. It uses the formalism of anomaly detection and extends it to spatio-temporal event analysis. Here, the \emph{spatial component} is a graph of interconnected entities (e.g. web pages, users or documents), and the \emph{temporal component} is the activity logs of these entities. A spatio-temporal event is a set of nodes that have an abnormal behavior during a period of time. Importantly, the nodes may not undergo a change at the same time. There can be a complex interaction between the nodes during an event when a change in one node can trigger changes in others. This complex spatio-temporal evolution is not covered in anomaly detection literature.

2) For each event or anomaly detected using our method, the model produces a rich set of spatio-temporal indicators rather than a single label or a probability of being an anomaly. In this sense, it is closer to the spatio-temporal data mining~\cite{atluri2018spatio}, where the purpose is to extract the anomalous events and keep as much information about them as possible for the sake of interpretability. Providing the description of detected anomalies to domain experts is a powerful feature of our data mining process. Our approach provides insights on the collective behavior in web and social networks, how the visitor activity evolves and propagates over the network. We illustrate that in the Experiment section and reveal new insights on the collective behavior of Wikipedia visitors.

3) We define the concept of \emph{potential anomaly} that introduces a prior on the presence of an anomaly and enhances the scalability of the method. Indeed, in many anomaly detection applications and thanks to expert knowledge, the data can be separated into 2 parts; one part contains potential anomalies, while the other contains non-anomalous samples. Our concept defines the separation in a rigorous manner and allows discarding non-anomalous samples. This step significantly reduces the amount of data to process. The concept of \emph{potential anomaly} is general and can be used for other methods and applications.



\begin{figure}[t!]
	\centering
	\includegraphics[width=\columnwidth, clip, trim={10.7cm, 6.9cm, 9.2cm, 6.5cm}]{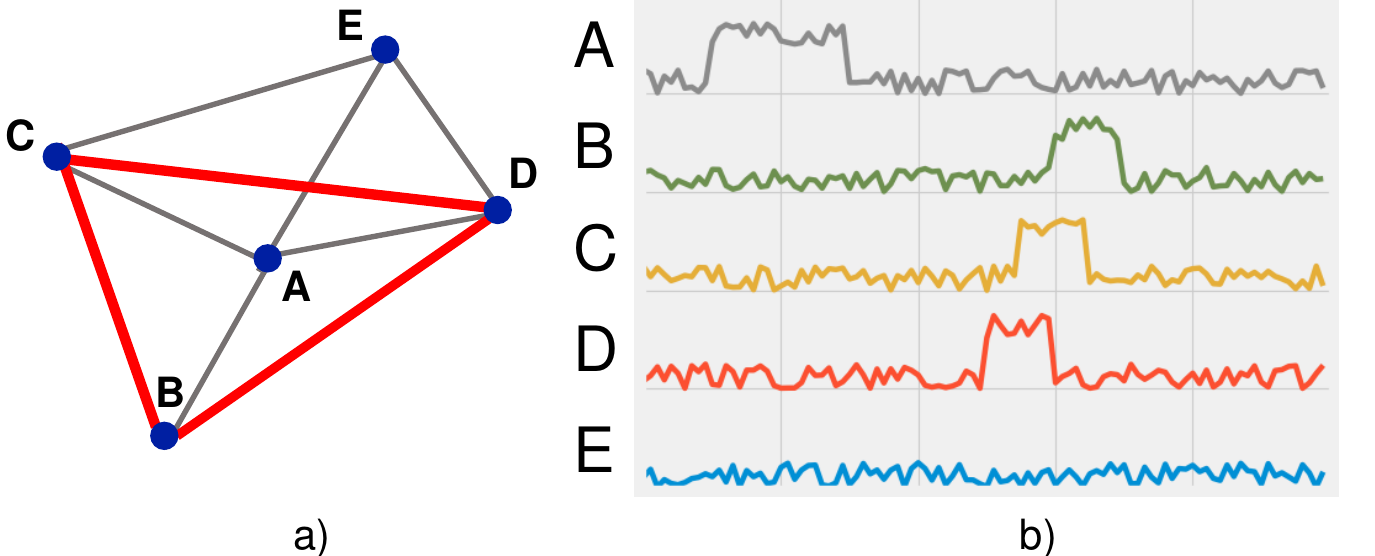}
	\caption{ Spatio-temporal data structure combining a graph topology and time-series. a) Graph topology. Edges highlighted in red depict the spatial component of an anomalous spatio-temporal pattern. b) Time-series signals residing on the vertices of the graph. Signals associated to nodes B, C and D are correlated: an anomalous process is propagating from node D to B through C. This is an abstract illustration of a dynamic anomaly detected by our method.}
	\label{fig: ts_graph}
\end{figure}

Our approach can be applied to the data that has a structure depicted in Fig.~\ref{fig: ts_graph}. It consists in an attributed graph where the attributes of the nodes are time-series signals.
It leverages the network structure of the given data to detect dynamic events or anomalies in collective behavior.

In the experiments, we use two spatio-temporal datasets. First, we build a network of Wikipedia articles and use visitor activity of each article, i.e. the number of visits per article per hour, as a node attribute. The static underlying network is the Wikipedia hyperlinks network. Two pages are connected if there is at least one hyperlink reference between them. Second, we transform the Enron email dataset into a temporal network where nodes correspond to email addresses of employees and edges reflect an email exchange between a couple of employees. The temporal activity attributes of the nodes is the number of emails sent per day by an employee.

In the results of our experiments, we demonstrate that we can extract anomalous patterns in collective behavior of users in web and social networks. The anomalies correspond to subgraphs containing nodes whose behavior deviates from the norm. For the Enron email dataset, these subgraphs are groups of employees having an increase in their email exchange during a short period of time corresponding to major events in the corporation. For the Wikipedia data, the subgraphs contain linked pages closely related to an event that triggered a sudden increase of visits during a short period of time. These clusters of anomalous nodes can then be used for more detailed investigation, as shown in Section~\ref{Experimental results}.



The strength of our approach is that it provides a comprehensive description of detected anomalies. As a result, we are able to perform a thorough qualitative evaluation of our results. Although, it turns into a difficulty when a quantitative evaluation of the method is needed. It is not as straightforward as computing the accuracy of a classification. Therefore, we use alternative methods for validating of the results. First, we use Google Trends as a ground truth indicator of the anomalous activity of the visitors on the Web. We verify the detected anomalous events using the trending topics extracted from Google Trends. In the second case, the Enron email dataset contains ground truth, hence we use it to validate the detected anomalies.

\section{Related work} \label{related work}


\noindent{\bf Anomaly detection}.   A recent review of the emerging field of spatio-temporal data mining highlights the importance of anomaly detection techniques for the dynamic networks domain~\cite{atluri2018spatio}. Due to the complex nature of the data, most existing approaches treat spatial and temporal components of the data independently~\cite{wu2010spatio, lu2007detecting, faghmous2013multiple}. There are several application-specific approaches for anomaly detection in video streams that use spatial and temporal information jointly~\cite{li2014review, kratz2009anomaly}. Lappas et al.~\cite{lappas2013stem} presented an approach for bursts identification in Twitter using spatial and temporal aspects of the data, although did not use the formalism of anomaly detection.

When dealing with the data from social networks, event detection is closely related to anomaly detection, therefore there exists a number of approaches that use spatio-temporal features of these datasets for collective behavior analysis. The category of Event detection in~\cite[Type~4]{ranshous2015anomaly} covers the case where all nodes of a subgraph contribute to the creation of an event at the same time. This is also the case in~\cite[Def. 4]{akoglu2015graph}, where the authors defined it as a problem of the dynamic-graph anomaly detection. In addition, these reviews as well as~\cite{aggarwal2014evolutionary},  focus on dynamic graphs, where the graph structure evolves over time, whereas in our case, the structure of the graph is static.

Our method is developed for static graphs with a dynamic evolution of node attributes (time-series). Our definition of an anomaly is more general since we track a heterogeneous spatio-temporal pattern that can emerge in situations when the anomaly spread or propagate over the network. It can not be captured by a single subgraph. Within our framework, an anomaly is described by two components. First, \emph{a graph pattern} that involves multiple nodes, possibly anomalous at different time steps. Second, \emph{a temporal pattern}, a set of time-series, that reflects the anomaly evolution on each node. Tracking these types of patterns has a number of applications~\cite{atluri2018spatio,benzi2016principal,griffa2017transient}, although they have not yet been linked to the field of anomaly detection.

\noindent{\bf Enron email dataset}. Several studies have investigated the Enron email dataset using different approaches. While many of them focus on email contents and perform analysis using natural language processing techniques, others focus on the network structure and anomaly detection. In~\cite{diesner2005communication}, the authors explore the email communication network of the corporation. Wang et al.~\cite{wang2014locality} adopt an anomaly detection framework to spot remarkable events. We compare our results to this latter reference to validate the accuracy of our detection.

\noindent{\bf Wikipedia dataset}. There is a large number of studies on mining the visitor or editor activity on Wikipedia that are aimed at getting better insights on collective behavior and social interactions~\cite{tinati2016finding}, \cite{garcia2017memory}, \cite{kanhabua2014triggers}. However, due to a large amount of data, the aforementioned studies are restricted to particular topics of interest and subsets of selected Wikipedia articles.
For instance, only traumatic events such as attacks and bombings have been investigated in \cite{ferron2011studying}, \cite{ferron2012collective} based on the Wikipedia edit activity data. Analyzing Wikipedia daily page views, Kanhabua et al.~\cite{kanhabua2014triggers} investigated 5500 events from 11 categories such as aviation accidents, earthquakes, hurricanes, or terrorist attacks. Wikipedia hourly visits on the pages of celebrities were used to investigate the fame levels of tennis players \cite{yucesoy2016untangling}. These studies point out the high interest in this dataset and the increasing need for more systematic detection methods.

Mongiovi et al.~\cite{mongiovi2013netspot,mongiovi2013mining} provided the first investigation from an anomaly detection point of view, where the Wikipedia page counts data are combined with the graph of hyperlinks. However, they apply their method to a pre-selected subset of Wikipedia. 

Due to the introduced concept of potential anomaly and the distributed implementation, our method can handle the full Wikipedia network and long-term visitor activity records.

\section{Method} \label{Graph learning}
Anomaly detection in dynamic networks generally contains two stages~\cite{ranshous2015anomaly}. The first stage is usually responsible for feature extraction from the domain-specific data. The second stage applies the anomaly detection algorithm to the extracted features and reports the detected anomalies.

We build our method upon this scheme in the following way. Firstly, we add to the Stage 1 a filtering step that keeps only potentially anomalous signals. It allows reducing the amount of data that is processed in Stage 2 by discarding obviously non-anomalous cases.
Secondly, in Stage 2, our model contains a learning step that enhances the interpretability of the detected anomalies. It provides a complete and detailed spatio-temporal description of the detected anomaly, rather than the classification \emph{anomalous/non-anomalous} commonly seen in anomaly detection. This description is a group of interconnected nodes (spatial information), where every node has a time-series attribute (temporal information) that indicates the time when the anomaly occurred.

The learning step is inspired by the model of a memory neural network, the Hopfield network with the Hebbian learning rule. In our model, we adapt the learning rule to fit our spatio-temporal data structure in the following way.
An edge weight is reinforced when two neighbors are active together during the same time slice. This particular network design shares similarities with the Hotspot anomaly detection for graph streams~\cite{yu2013anomalous}, however, the authors use it for feature engineering (Stage 1 of the processing), while in our approach this update is a part of Stage 2, the learning and classification stage.

The Hopfield network approach learns a memory network, where the nodes correspond to the ones of the initial graph. During the learning process, edges between the nodes will be either strengthened or removed depending on the temporal behavior of each node. As a result of learning, nodes with similar behavior are connected by strong links and clustered together in the memory network. These clusters contain groups of nodes with common abnormal behavior.



\noindent\textbf{Stage 1: Feature extraction and filtering.} 
This stage highlights anomalies and extracts features from the raw data. We introduce a notion of \emph{potential anomaly} and we filter out the nodes that are not potentially anomalous.
Let $V$ be the set of vertices of graph $G$ and  $x_i[t]\in \mathbb{R}$ be the value associated to vertex $v_i\in V$ at the time $t\in [0,T-1]$. The time-series have a length of $T$ samples.
\begin{definition}[Single vertex potential anomaly]
Given a score function $f_i:\mathbb{R} \to \mathbb{R}$ for each vertex $v_i$ and a threshold value $c_0\ge 0$, a vertex $v_i\in V$ is said to have a potential anomaly at time $t$ when its time series value is such that $|f_i(x_i[t])|>c_0$.
\end{definition}
Note that the potential anomaly is local on the graph, i.e. it depends only on the time-series attribute of a considered vertex. If the score function is such that $f_i=f-\hat{f}(v_i)$, for some function $f$ and $\hat{f}(v_i)$ a summary statistics of the scores $f(v_i[t])$, our definition corresponds to the definition of the anomaly in~\cite{ranshous2015anomaly}.
A basic example of a potential anomaly is the time-series values exceeding a fixed threshold $c_0$. In that case, $f_i$ is the identity. 

In this work, potential anomalies have to satisfy the following requirement. A potentially anomalous node must have a sufficient number of bursts in their time-series attribute. Unless this requirement holds, we discard the node. To define a burst of activity we compute the mean $\mu_i$ and the standard deviation $\sigma_i$ of $x_i$ over time. We normalize the time-series by defining the score function $f_i^b(x_i[t])=(x_i[t]-\mu_i)/\sigma_i$. We select values that are above $c_0^b$, the activity rate parameter ($c_0^b=5$ for Wikipedia, $c_0^b=3$ for Enron). The \emph{burstiness} $b_i$ of a signal $x_i$ of a node $i$ is
\begin{equation} \label{eq: burst}
b_i = \sum_{t = 0}^{T-1}k_i[t], \qquad
k_i[t] = \begin{cases}
1, &\text{if $|f_i^b(x_i[t])| > c_0^b$},\\
0, &\text{otherwise}.
\end{cases}
\end{equation}
The minimal number of bursts (potential anomalies) per vertex in Wikipedia is $b_i=5$ while in the email dataset it is $b_i=2$.

The general definition of the potential anomaly allows for other score functions, such as a moving average or ARMA filter prior to the thresholding, a short time Fourier transform or a wavelet transform. This depends on the dataset and should be chosen carefully to make a compromise between efficiency and scalability.

Applying $f_i$ to the time-series gives us 1) features reflecting an anomaly and 2) an initial indication of anomalous behavior. In practice, removing nodes with no potential anomaly reduces the amount of data by an order of magnitude without losing relevant information that is required for the anomaly detection.





\noindent\textbf{Stage 2: Hopfield network learning}. The presented anomaly detection approach is aimed at detecting groups of vertices that have a similar, abnormal behavior. The learning stage is intended to make this coherent behavior apparent in the memory network. 

Our approach is based on the Hopfield model of artificial memory~\cite{hopfield1982neural}. It is an unsupervised learning method.
To implement it, we use a synaptic-plasticity-inspired computational model, the Hebbian learning rule~\cite{hebb2005organization}. The main idea of this model of brain memory is that a co-activation of two neurons results in the reinforcement of a connection (synapse) between them. Although, contrary to the original learning rule, in our model, we do not take causality of activations into account. 

In our case, the Hopfield network has $N$ nodes. These nodes correspond to the ones given in the dataset, after the reduction performed at Stage 1 (only the nodes containing potential anomalies). 

The learning process is as follows. We use the initial structure of the given network. For two initially connected nodes $i$ and $j$ of the Hopfield network, at time $t$, we update the weight of an edge $e_{ij}$ between them according to the similarity measure ${{\rm Sim}\{i,j,t\}}$. Note that we only perform this step for the nodes that are initially connected and do not compare every possible pair of nodes. This is crucial for the tractability of the method in cases when we deal with large networks.
For each time step $t$, the edge weight $w_{ij}$ between $i$ and $j$ is updated as follows: 
\begin{equation} \label{w_update}
\Delta w_{ij} = \begin{cases}
+Sim\{i,j\}, &\text{if $Sim\{i,j\}$ $>$ $\lambda$},\\
-\alpha Sim\{i,j\}, &\text{otherwise},
\end{cases}
\end{equation}
where $\lambda\ge0$ is the \emph{sparsity parameter}. Similarly to firing neurons, nodes expressing similar behavior have their connection weight increased. When $\alpha>0$, the weight decreases allowing older events to be forgotten in order to keep only the latest anomalies. In our experiments, we set $\alpha=0$, as we want to keep all the anomalies. We also fix $\lambda=0.5$. The value of $\lambda$ influences the sparsity of the final network. Increasing $\lambda$ reduces the number of edges in the resulting memory network. One should increase the value of $\lambda$, when looking for the most outstanding anomalies and decrease it, when a higher sensitivity is required.

Before describing the similarity function, let us introduce the activity function $y_i$ at a node $v_i$:
\begin{equation}\label{eq:activity_function}
y_i[t]=x_i[t]\times k_i[t].
\end{equation}
Note that the activity function can also be defined from the features calculated at Stage 1, $y_i[t]=f_i(x_i[t])\times k_i[t]$. In the present work, we use the definition~\eqref{eq:activity_function}.

Different similarity measures can be defined. A first example is the following:
\begin{equation} \label{similarityPearson}
{\rm Sim}\{i,j,t\} = y_i[t]y_j[t].
\end{equation} 
When $y_i$ and $y_j$ are normalized, this measure gives the Pearson correlation between the nodes (if $\lambda=0$ and $\alpha=1$). The $L^2$ distance with a Gaussian kernel
${\rm Sim}\{i,j,t\} = \exp(-|y_i[t]-y_j[t]|^2)$ can also be a good candidate.
Another measure, restricted to the interval $[0,1]$ and that worked well in our applications, is defined as follows: ${{\rm Sim}\{i,j,t\}}=0$ if $ y_i[t]=y_j[t]=0$ and otherwise,
\begin{equation} \label{similarity}
{\rm Sim}\{i,j,t\} =
\frac{\min(y_i[t], y_j[t])}{\max(y_i[t], y_j[t])}\in[0,1].
\end{equation} 

\noindent\textbf{Recalling process}. Starting from an initial partial memory pattern $P_0\in \mathbb{R}^{N\times T}$, the recall of a learned pattern is done by the following iterative computation:

\begin{equation} \label{eq: hopfield}
P_{j+1} = h_{\theta}(WP_{j}),
\end{equation}
where $W\in \mathbb{R}^{N\times N}$ is the weight matrix of the Hopfield network. The function $h_{\theta}:\mathbb{R}^{N\times T}\to \mathbb{R}^{N\times T}$ is a nonlinear thresholding function (step function giving values $\{-1,1\}$) that binarize the vector entries. The value $\theta$ is the threshold (same for all entries). In our case, we build a network per month so $W$ is associated to a particular month. For each $j\ge0$, $P_j$ is a matrix of binarized time-series where each row is associated to a node of the network and each column corresponds to a time step of the month considered. We stop the iteration when the iterative process has converged to a stable solution ($\|P_{j+1}-P_j\|\le\varepsilon$, where $\varepsilon$ is small, the norm is the Frobenius norm). The initial pattern $P_0$ is a binary matrix, where the rows have all values set to $-1$ (inactivity) except the ones associated to the partial memory pattern obtained from the time-series using the expression of $k_i$ defined in Eq.~\eqref{eq: burst}.
The computation of the iterative process is efficient as the matrices are sparse and in practice, it converges after a reasonable number of steps.


\noindent\textbf{Graph visualization and community detection}. Since the connections between nodes with similar activity are reinforced during the learning process, after we prune low-weight edges, each Hopfield network transforms into a modular graph structure with strongly connected clusters of nodes having a similar activity. These groups can be either isolated connected components or communities within the largest connected component. The analysis of the Hopfield networks and their communities provides a good way to spot, analyze, and interpret the anomalies in the dynamics as we will see in the next section.
To find communities, we use the Louvain method \cite{blondel2008fast} and to represent the graph in 2D space for visualization, we use a force-directed layout \cite{jacomy2014forceatlas2}. Additional and more interactive visualizations are available online \cite{WikiViz}.

\noindent\textbf{Storing memories of detected anomalies}. The number of anomalies that can be memorized by a single network is limited~\cite{mceliece1987capacity}. Indeed, without the forgetting parameter $\alpha$, the clusters of nodes will accumulate inside the graph, eventually overlapping and forming larger clusters of unrelated anomalies. To avoid this but still keep track of older events, we create snapshots of memories by slicing the time-series into time-windows of the finite duration. The time-window size depends on the application. For example, in the analysis of Wikipedia data, we use time-windows of one-month length.
 
\noindent\textbf{Complexity}. The computations are tractable because 1) they are local on the graph, i.e. weight updates depend on a node and its one-hop neighbors, 2) weight updates are iterative, and 3) a weight update occurs only between initially connected nodes and not among all possible combinations of nodes. These three facts allow us to build a distributed model to speed up computations. For this purpose, we use a graph-parallel Pregel-like abstraction, implemented in the GraphX framework \cite{gonzalez2014graphx}, \cite{xin2013graphx}.

Stage 1 has complexity $\mathcal{O}(NT)$ where $N$ is the number of nodes and $T$ the number of time steps. Stage 2 has a complexity of $\mathcal{O}(ET)$, where $E$ is the number of edges. The recall process involves a multiplication by a sparse matrix with $2E$ non-zero entries, hence the complexity is $\mathcal{O}(ET)$. It is important to point out that the number of nodes and edges in the computations are not necessarily the numbers found in the dataset: Stage 1 may discard a large number of inactive nodes (containing no potential anomaly), Stage 2 sparsifies the time series as Eq.~\eqref{eq:activity_function} may set to zero a large number of values.



\section{Datasets} \label{Dataset}

{\bf Wikipedia dataset.} We use the dataset described in \cite{benzi2017recommender}.
This dataset is based on two Wikipedia SQL dumps: English language articles and user visit counts per page per hour. The original datasets are publicly available on the Wikimedia website~\cite{wikimediaPageCounts}. 

The Wikipedia network of pages is first constructed using the data from article dumps that contain information about the references (edges) between the pages (nodes)\footnote{Note that Wikipedia is continuously updating. Some links that existed at the moment we made the dump may have been removed from current versions of the pages. To check consistency with past versions, one can use the dedicated search tool at \url{http://wikipedia.ramselehof.de/wikiblame.php}.}. Time-series are then associated to each node (Fig. \ref{fig: ts_graph}), corresponding to the visits history from 02:00, 23 September 2014 to 23:00, 30 April 2015. The time-series have a length of $T=5278$ hours.

{\bf Enron dataset.} We use Enron email dataset\footnote{https://www.kaggle.com/wcukierski/enron-email-dataset}. It contains 614586 emails sent over the period from 6 January 1998 until 4 February 2004. We remove the periods of low activity and keep the emails from 1 January 1999 until 31 July 2002 which is 1448 days of email records in total. We remove inactive email addresses that sent less than three emails over that period. In total, the Enron email network contains 6 600 nodes and 50 897 edges.


\section{Experiments and results}\label{Experimental results}
In this section, we present our results of the anomaly detection experiments. We apply the proposed algorithm to the two datasets. The results obtained from the Enron one serve mainly to validate the method so we do not describe the analysis of the data. We dedicate more space to the detailed inspection of the Wikipedia dataset, which is a larger and more informative dataset allowing us to demonstrate the scalability of our method and to provide the detailed interpretation of the results.

\begin{figure}[t!]
	\centering
	\includegraphics[width=\columnwidth, trim={3.5cm 1.5cm 4cm 1.5cm}, clip]{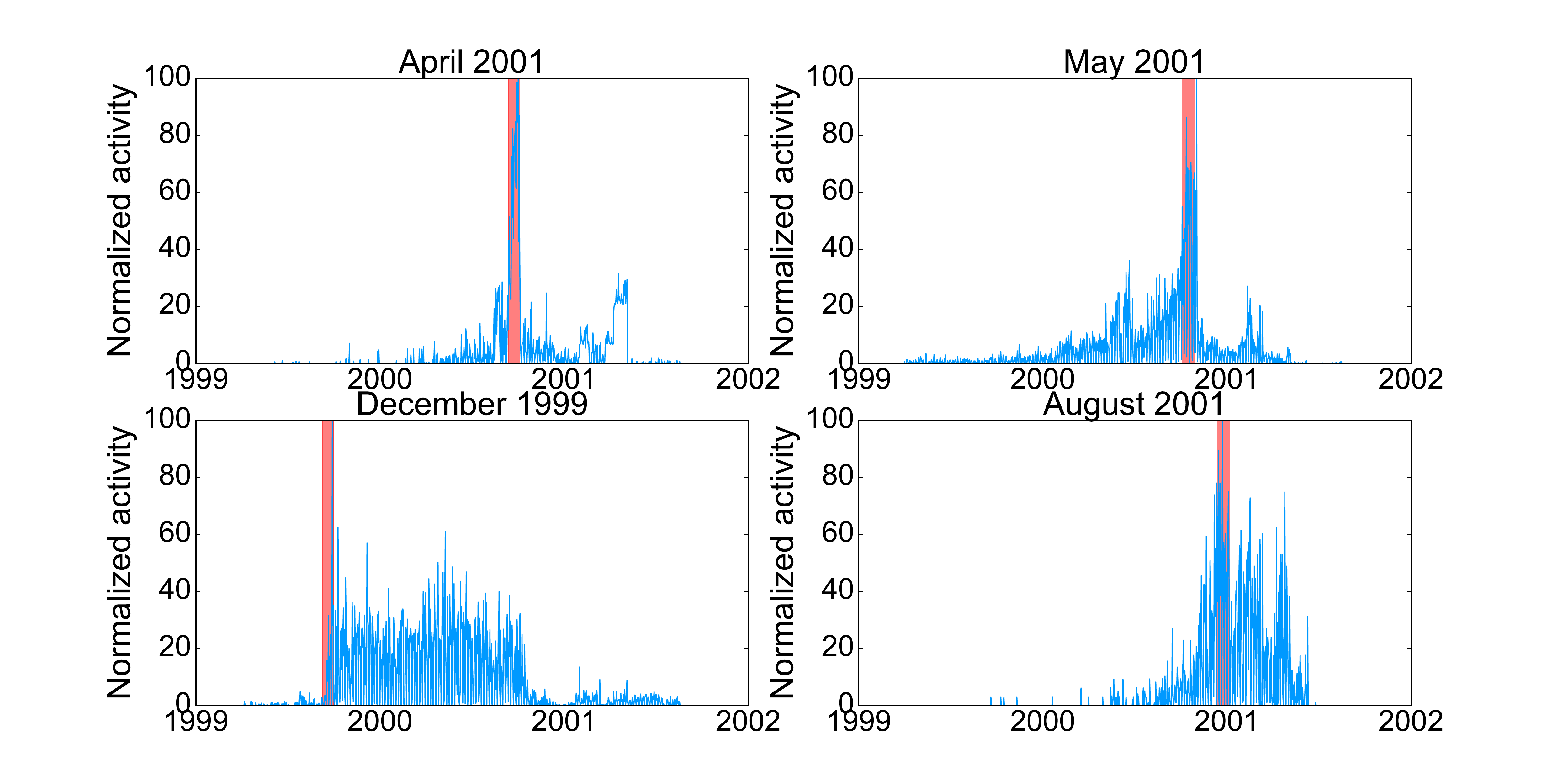}
	\caption{Anomaly detection in Enron email network. Red areas highlight the month periods previously reported as anomalies (ground truth supported by the real world events). Blue lines reflect the normalized (scale 0-100) overall activity level in the network computed by the proposed algorithm. We can see that the algorithm detects anomalies in all reported cases.}
	\label{fig: enron}
\end{figure}

\noindent{\bf Enron email dataset.} In order to apply our anomaly detection algorithm to this dataset, we represent the data as a time-series graph data structure, which is depicted in Fig.~\ref{fig: ts_graph}. 
The static graph is the network of email communications: the nodes are email addresses and they are connected if they have exchanged at least one email over the 5 years. It is an undirected, unweighted graph. Time-series associated to the nodes are captured from the email activity; each temporal value is the number of emails sent from the associated address during one day. Here, the connection between email addresses is reinforced if a similar amount of emails have been sent by both of them over the same hour.
Note that the graph and time-series design could have been more complex with directed edges or time-series associated to the edges. However, in order to stay as general as possible, we decided to strictly follow the steps described in Sec.~\ref{Graph learning}. 


In this case, an anomaly is a sudden increase in email communications among a group of employees of the corporation. After learning a Hopfield network for every month, we investigate their structure. We select four months where the results can be compared to the literature.
The authors of~\cite{wang2014locality} observe four anomalous periods and relate them to the specific news reports involving Enron. We use these events as a ground truth. These are the months of December 1999, April, May and August 2001.
For each month, we select the largest connected component of the learned graph. We sum up the activity of its nodes to get a single time-series representative of the group activity. We show this normalized activity on Figure~\ref{fig: enron}. We define an anomaly to be a spike of overall activity in a cluster of email addresses that we detect after learning. All the four curves have a larger activity during the chosen month than in the rest of the time span. For April and May 2001, it is more than twice the maximal activity for the rest of the months, showing the evidence of an anomaly. December 1999 and August 2001 increases in activity have a longer trace after the main anomaly has been detected but still, the high activity does not spread for more than one year. Concerning the monthly components of active nodes for the 4 chosen months, it involves 29, 25, 126, 28 nodes respectively for December 1999, April, May and August 2001. Almost all of them correspond to addresses of Enron employees. Except for the May event, the activity involves less than 30 persons. In May, it involves 100 employees out of the 158, showing an event that impacted the whole company.


The application of our method to the Enron dataset allows detecting all the anomalies presented in the state-of-the-art literature. It reveals the days of the peak activities, the duration of the events, and the persons involved, which makes more detailed investigation possible in the future. Besides, our approach can detect periodic or recurrent anomalies due to associative nature of the Hopfield network. This feature will be demonstrated in the following experiments on the Wikipedia data.

\noindent{\bf Wikipedia dataset.}
We conduct more detailed experiments on the Wikipedia web network and use Google Trends data as a ground truth to verify the accuracy of our findings. An anomaly, in this case, is a sudden increase in visitor activity in a small, local part of the web network. We start by analyzing the initial graph of Wikipedia web pages connected with hyperlinks. In this experiment, the time-series attributes of the nodes correspond to the viewership statistics of the associated web pages, described in Section~\ref{Dataset}.



 

To learn the global anomalies that occurred in the long-term dynamics of the Wikipedia web network, we build the Hopfield network of the 7-month period.
After the learning, only 275 498 edges have strictly positive weights (4.2\% of the initial graph). We remove the disconnected nodes and preserve only the largest connected component of the graph. The number of remaining nodes is 35 839 (31\% of the initial number).

\begin{figure}[t!]
	\centering
	\includegraphics[width=\columnwidth]{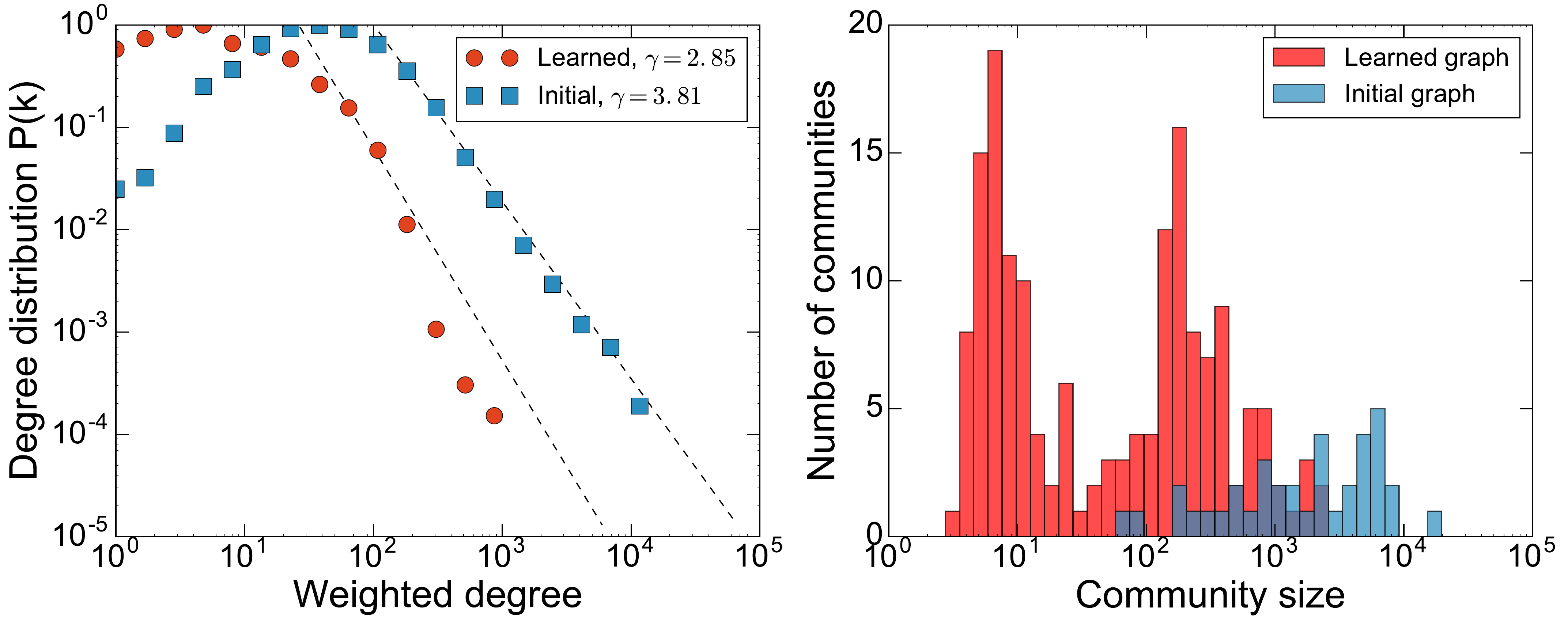}
	\caption{\textit{Left.} Weighted degree distribution in log-log scale for the Wikipedia graph and Hopfield network learned over the entire 7 months time span. Linearity in log-log scale corresponds to power-law behavior $P(k) \sim k^{-\gamma}$. The learned graph preserves a similar scale-free behavior, but is less connected and has fewer hubs than the initial graph. 
		\textit{Right.} Community size distribution of the initial Wikipedia graph of hyperlinks (blue) and the learned Hopfield network (red). The total number of communities: 32 for the initial graph, 172 for the learned one.}
	\label{fig: degree_log_log}
\end{figure}

The analysis of the static underlying graphs shows that both (the initial and learned) Wikipedia graphs have statistically heterogeneous connectivity (Fig.~\ref{fig: degree_log_log}).
However, the initial Wikipedia graph is dominated by large hubs that attract most of the connections to numerous low-degree neighbors. These hubs correspond to general topics in the Wikipedia network. They often link broad topics such as, for instance, the ``United States'' page, with a large number of hyperlinks pointing to it from highly diverse subjects. If we look at the viewership statistics, the activity of the visitors in these large clusters is uniform and does not expose any anomalies over time. We aim at extracting smaller communities that correspond to localized anomalies in the dynamics of the network. This is why we need the learned graph.

\begin{figure}
	\centering
	\begin{tabular}{cc}
		\includegraphics[width=0.45\columnwidth]{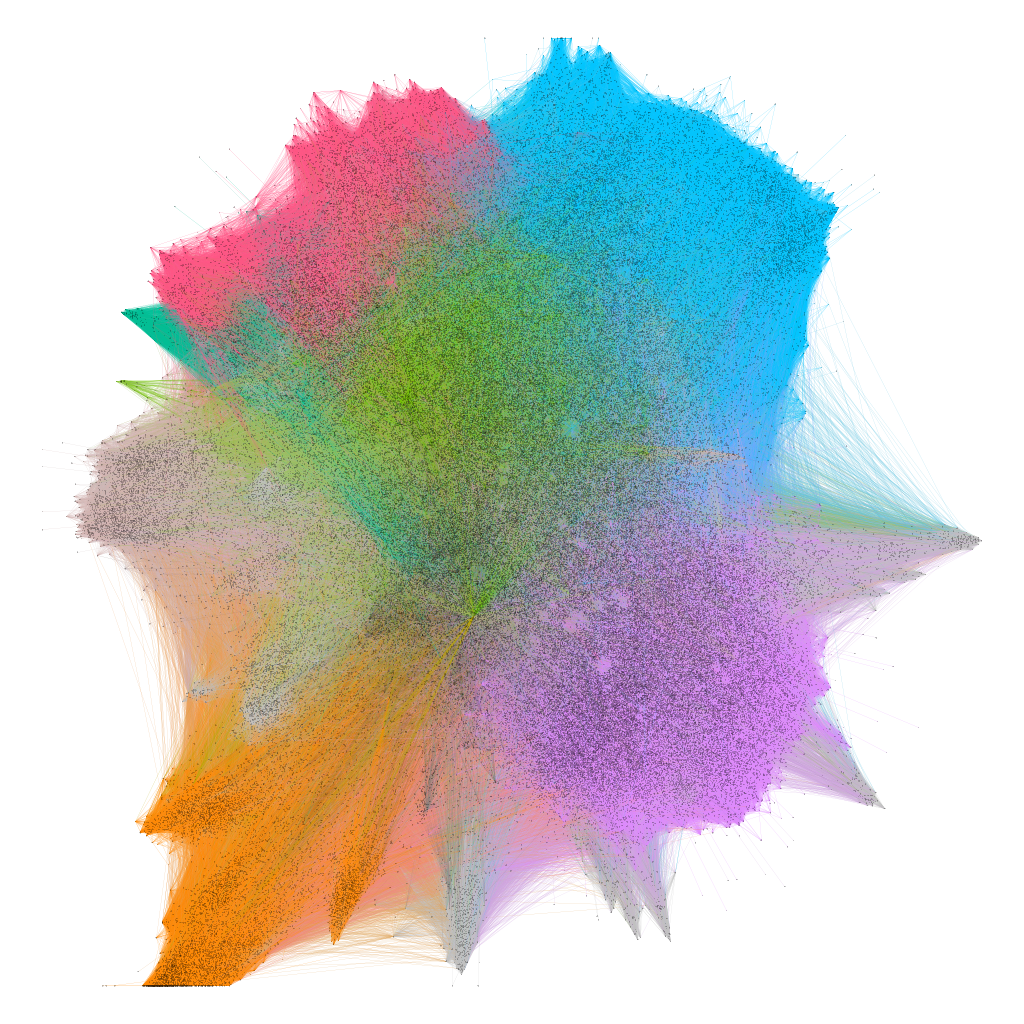} &
		\includegraphics[width=0.45\columnwidth]{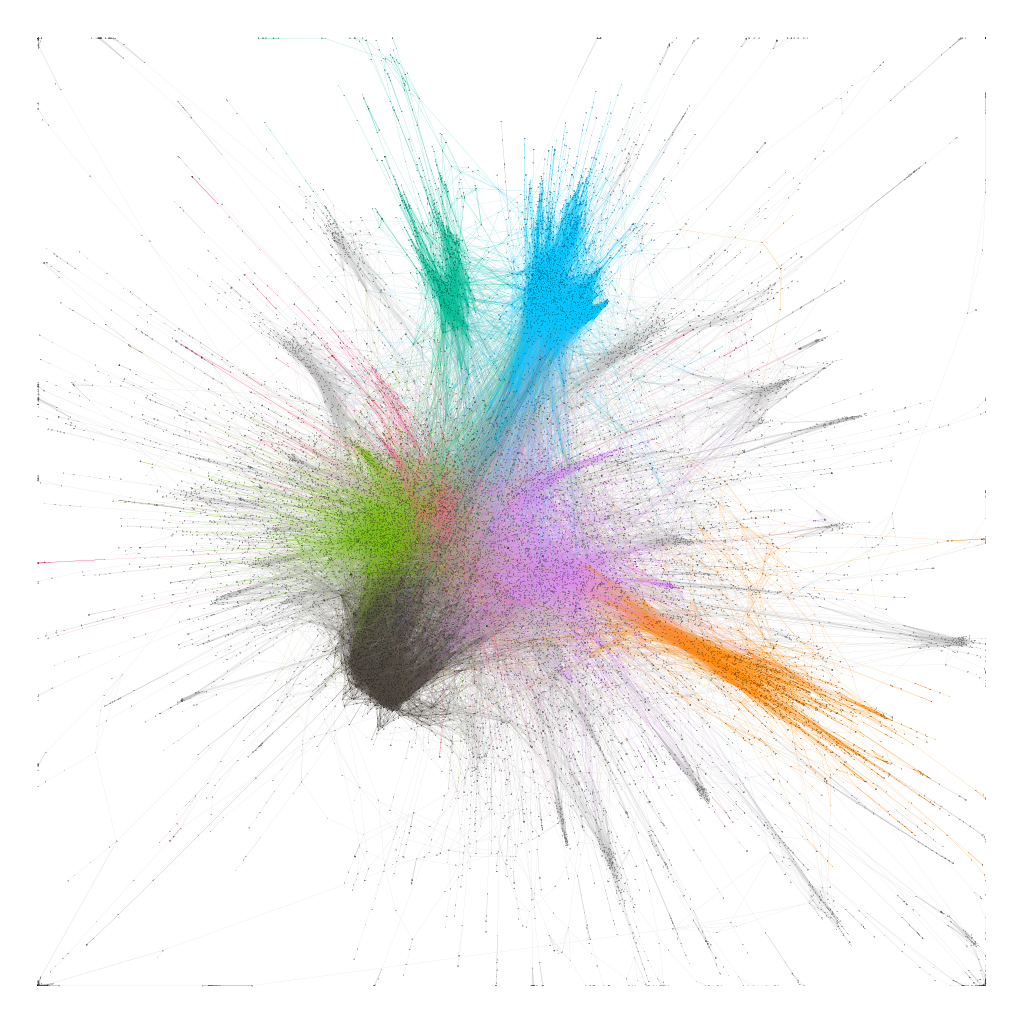}
		\\
		(a) Initial & (b) Learned \\[6pt]
	\end{tabular}
	\caption{Wikipedia graph of hyperlinks (left) and learned Hopfield network (right). Colors correspond to the detected communities. The learned graph is much more modular than the initial one, with a larger number of smaller communities. The layout is force-directed.}
	\label{fig: wiki_graphs}
\end{figure}

The visualizations of the initial and learned graphs using a force layout algorithm show striking differences (Fig.~\ref{fig: wiki_graphs}). The initial Wikipedia graph is dense and cluttered with a significant number of unused references, while the learned graph reveals smaller and more separated communities. This is also confirmed by the numerical measures such as the community size and degree distributions of the graphs (Fig.~\ref{fig: degree_log_log}, right). The number of communities and their size change after learning. Initially, the small number of large communities dominate the graph (blue), while after the learning (red) we see a five times increase in the number of communities. Moreover, as a result of the learning, the size of the communities decreases by one order of magnitude. The modularity of the learned graph is 25\% higher, strengthening the evidence of the creation of associative structures. These measures indicate that as a result of learning, we obtain a graph structure with small strongly connected clusters that correspond to a summary of anomalies in the network dynamics. We provide detailed examples of such anomalies in the following experiments.

The analysis of each community of nodes in the learned graph gives a glimpse of the events that occurred during the 7-month period and caused the anomalous behavior of visitors during that period of time. Each cluster is a group of pages related to a common topic such as a championship, a tournament, an awards ceremony, a world-level contest, an attack, an incident, or popular festive events such as Halloween or Christmas. 


Before going deeper into the clusters analysis and the corresponding anomalies in the network dynamics, we investigate the evolution of the graph structure over time with an emblematic example.

\noindent\textbf{Monthly networks.} As stated in Sec.~\ref{Graph learning}, we split the dataset into one-month periods. These periods are longer than the duration of an average event attracting the attention of Wikipedia users, that usually lasts no longer than two weeks. This means that we are going to detect multiple anomalies at once. Monthly graphs are smaller, compared to the 7-months graph, and contain 10 000 nodes on average. However, the properties and distributions of monthly graphs are similar to the 7-months one, described above. 


\begin{figure*}[t!]
	\includegraphics[width=\textwidth]{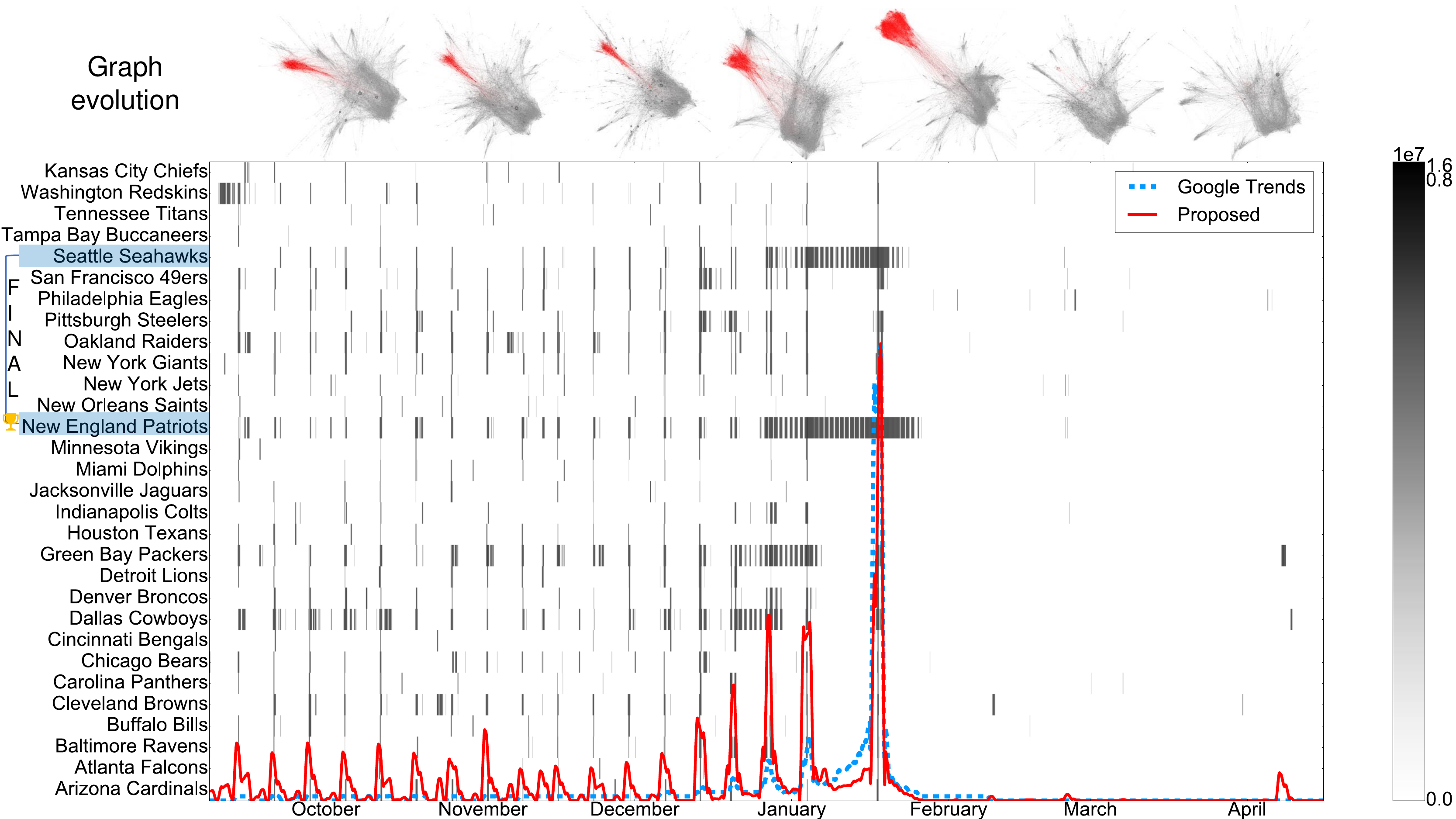}
	\caption{Evolution of the National Football League 2014-2015 championship cluster and visits on its articles. We show 30 NFL teams from the main cluster. Top: the 7 monthly learned graphs in gray, with the NFL cluster highlighted in red. The month is displayed on the bottom of the figure. Middle table: visitors activity per hour on the NFL teams' Wikipedia pages in greyscale (the more visits, the darker). Bottom: the red and blue curves are the total number of visits in the articles of the clusters over time (normalized) and the Google Trends curve for the keyword "Super Bowl", respectively. }
	\label{fig: SB_evolution}
\end{figure*}

To give an example of a detected event and the associated anomalous behavior of the Wikipedia users, we discuss the cluster of the USA National Football League championship, spotted in and extracted from the Hopfield network during several months between 2014 and 2015 (Fig.~\ref{fig: SB_evolution}). 

NFL is one of the most popular sports leagues in the USA and it triggers a lot of interest to the related articles on Wikipedia. Due to the high number of visits on this topic we were able to localize a cluster related to the NFL on each of the monthly graphs. Figure~\ref{fig: SB_evolution} shows the detailed information about the NFL clusters. The top part of the figure contains the learned graphs for each month, where the NFL cluster is highlighted in red.

The final game of the 2014 season, Super Bowl XLIX, had been played on February 1, 2015. This explains the continuous expansion of the cluster until February where its size reaches the maximum. The activity collapses right after this event and the cluster disappears.

For the sake of figure interpretability, we extracted 30 NFL team pages from the original cluster (485 pages) to show the details of the evolution in time as a table on Fig.~\ref{fig: SB_evolution}. This fraction of the nodes reflects overall dynamics in the entire cluster. Each row describes the hourly activity of a page, while the columns split the plot into months.
The sum of visits for the selected pages is plotted as a red line in the bottom.

The dynamics of the detected cluster reflects the real timeline of the NFL championship. The spiking nature of the overall activity corresponds to weekends when most of the games were played. Closer to the end of the championship, the peaks become stronger, following the increasing interest of fans. We see the highest splash of the activity on 1 February, when the final game was played.

We want to emphasize that this cluster, as well as all the others, was obtained in a completely unsupervised manner. The football teams pages were automatically connected together in a cluster having ``Super Bowl'' as a common topic. Moreover, the cluster is not formed by one Wikipedia page and its direct neighbors, it involves many pages with distances of several hops in the graph. 


The NFL championship case is an example of a periodic (yearly) event. The interest increases over the months until the expected final event, which causes an anomaly in the network dynamics. Accidents and incidents are the events of a different nature as they appear suddenly, without prior activity. The proposed learning method allows detecting this kind of events and the related anomalies as well. We provide examples of three accidents to demonstrate the ability of our method to detect anomalous behavior in the network in case of an unexpected event.

\begin{figure*}
	\begin{tabular}{ccc}
		\includegraphics[width=0.315\textwidth, trim={0cm 8.5cm 0cm 8.5cm}, clip]{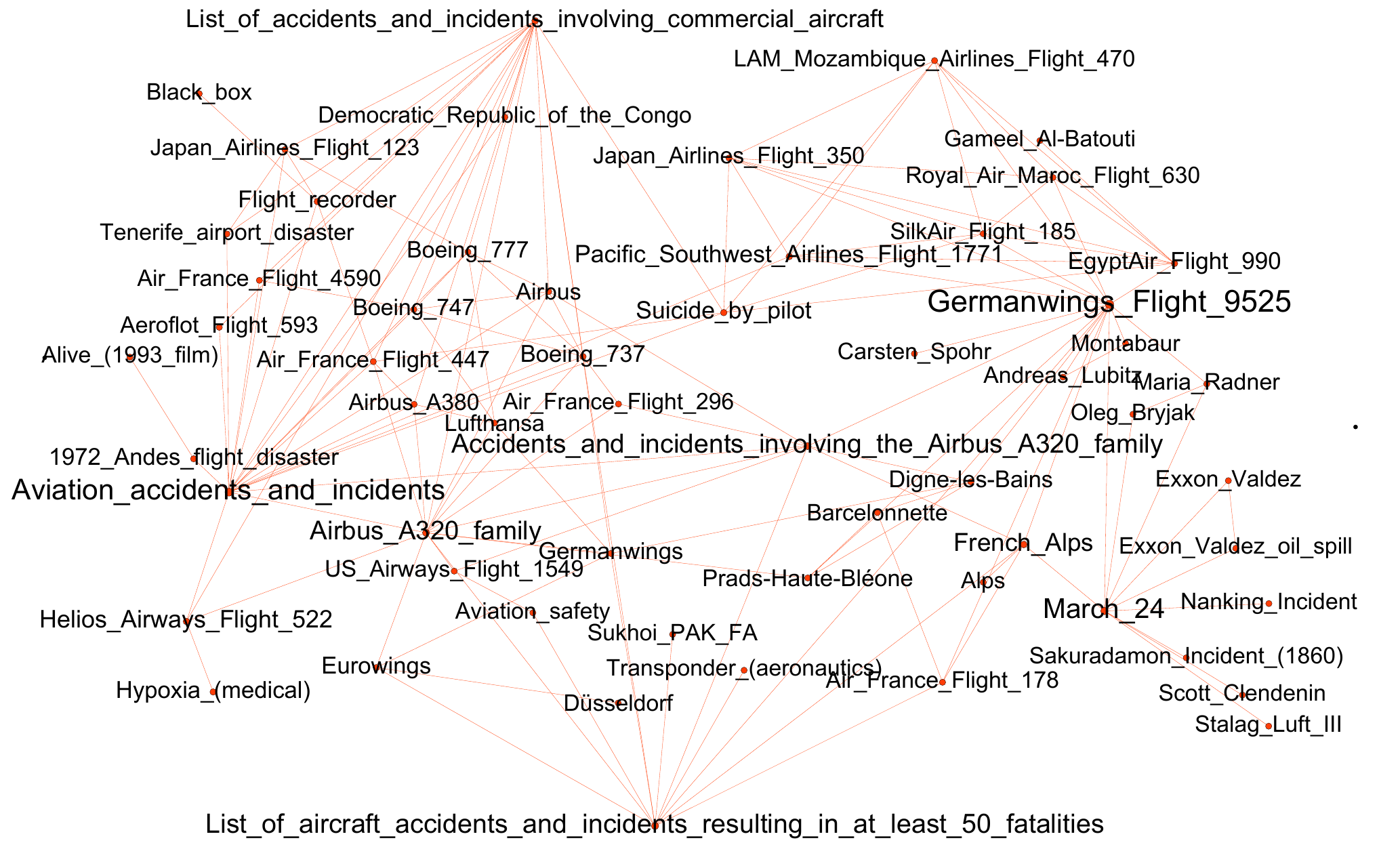} &
		\includegraphics[width=0.315\textwidth, trim={0cm 8.5cm 0cm 8cm}, clip] {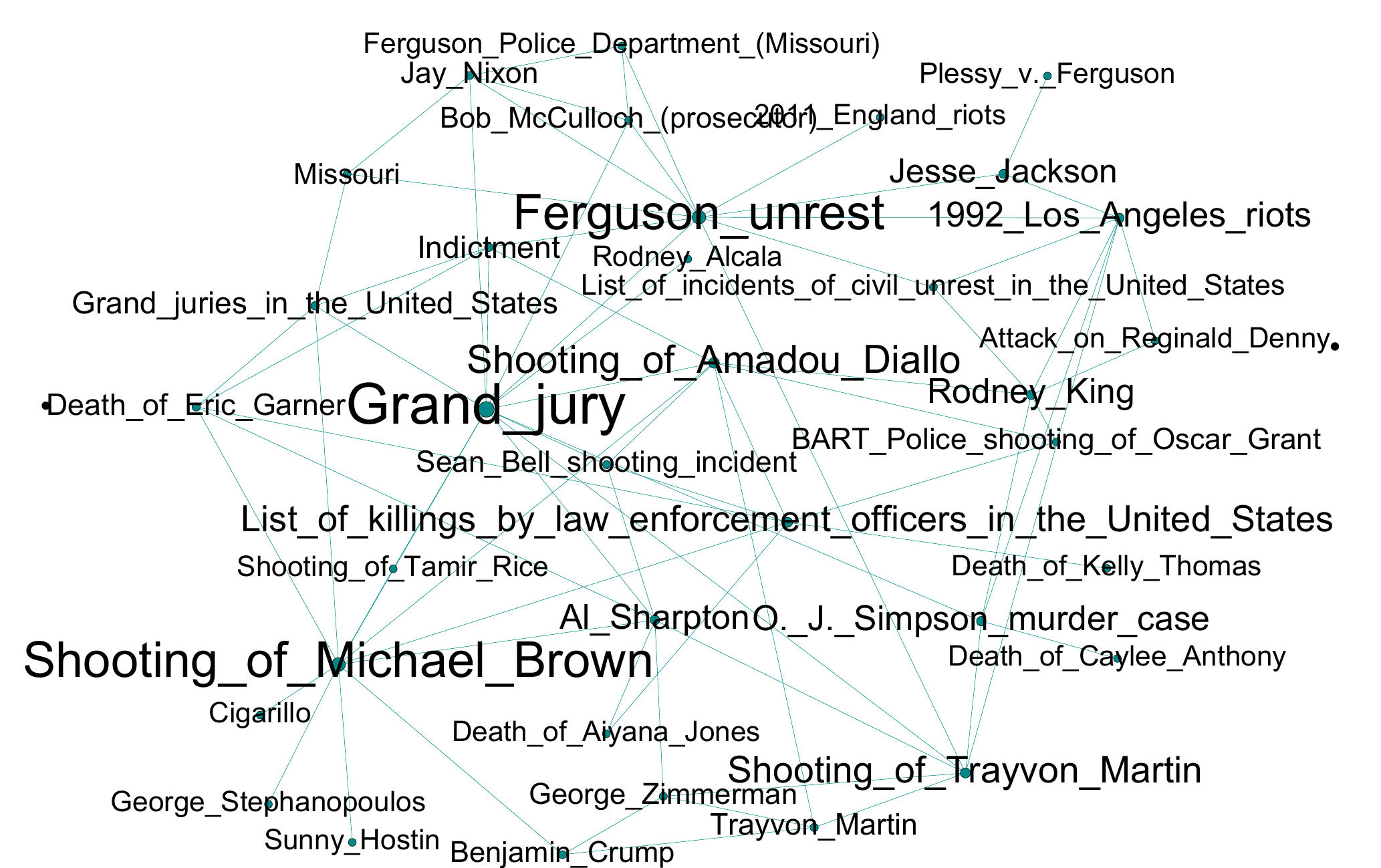} &
		\includegraphics[width=0.315\textwidth, trim={0cm 8.5cm 0cm 8.5cm}, clip]{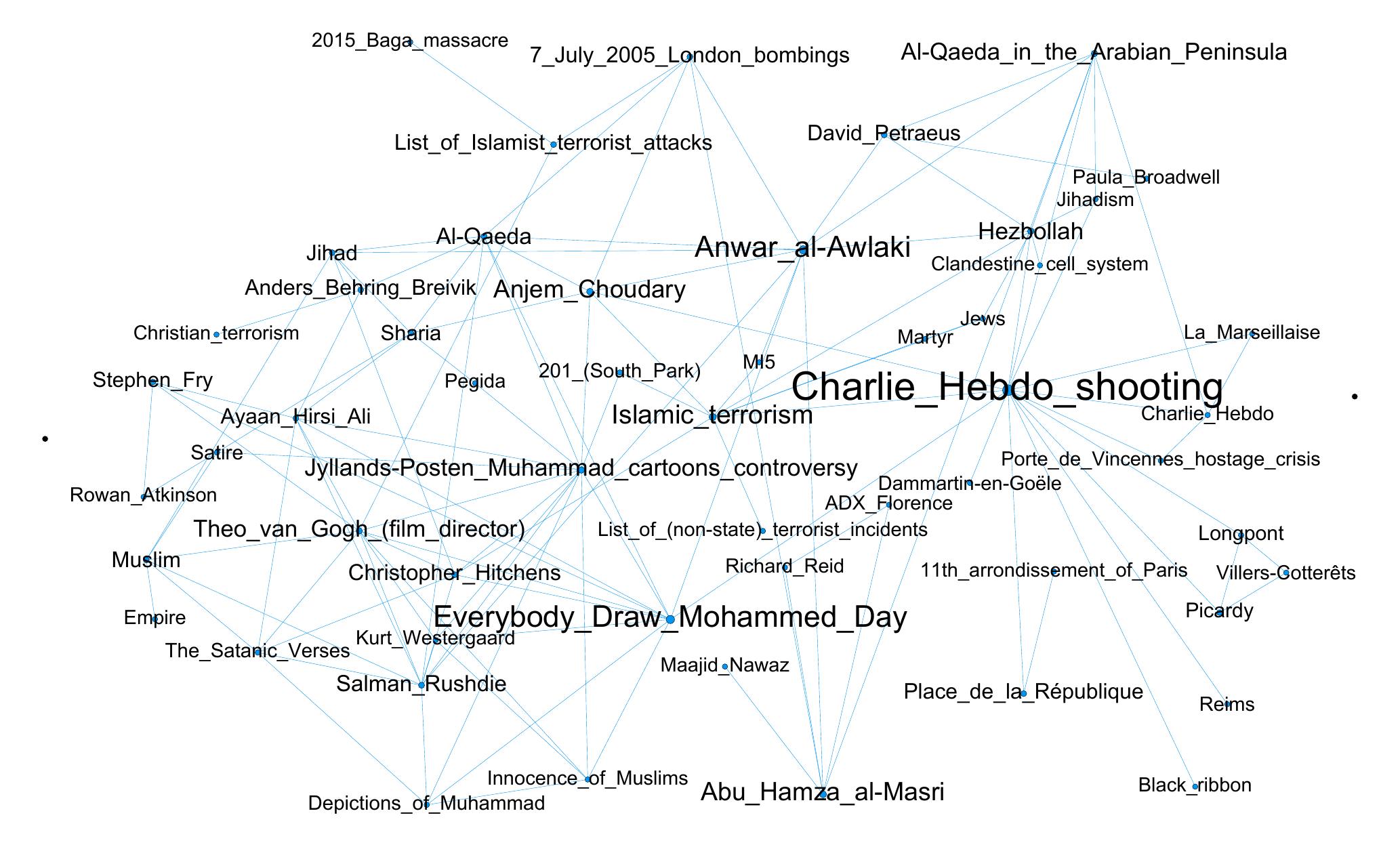}\\
		\includegraphics[width=0.315\textwidth,  trim={5cm 0cm 5cm 0cm}, clip] {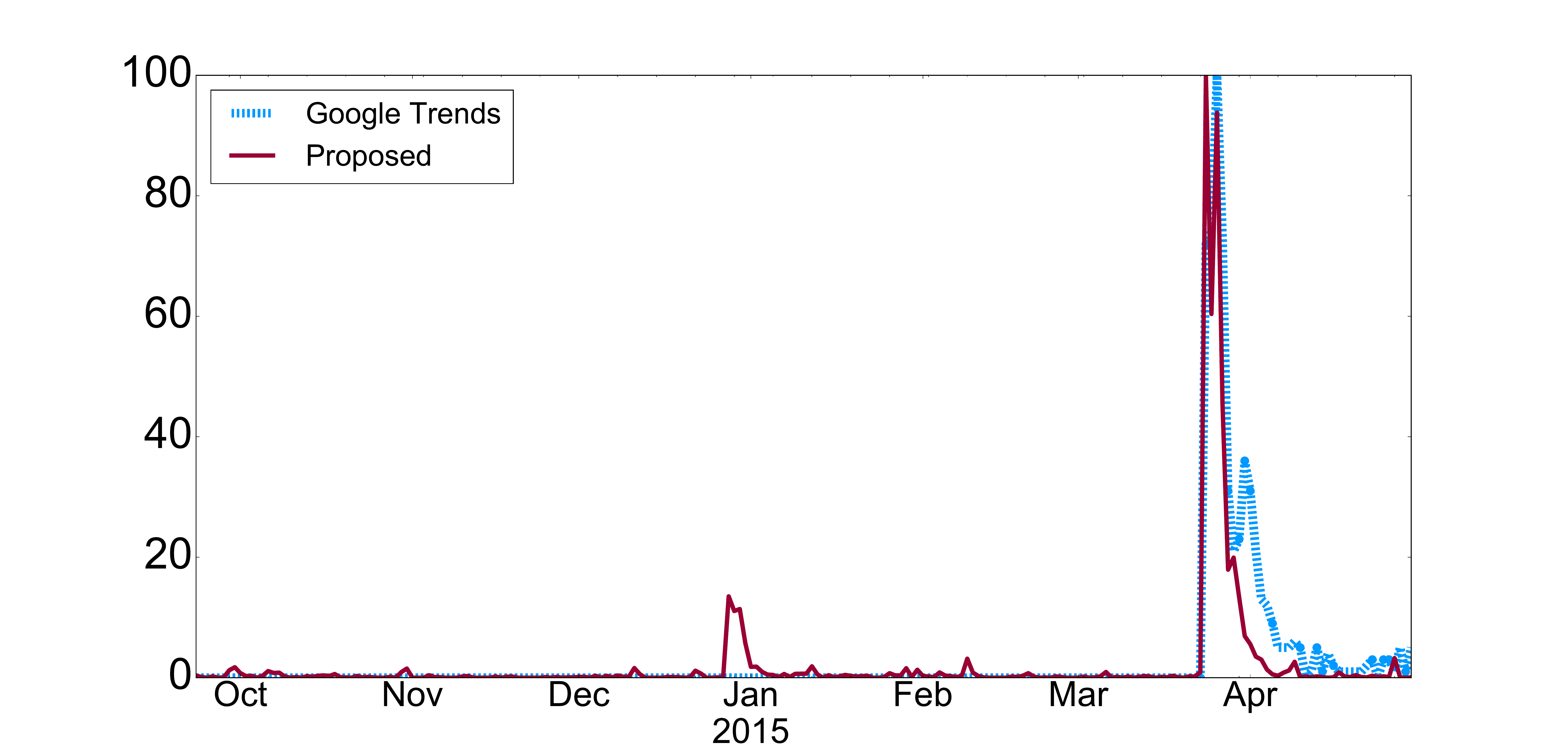} &
		\includegraphics[width=0.315\textwidth,  trim={5cm 0cm 5cm 0cm}, clip] {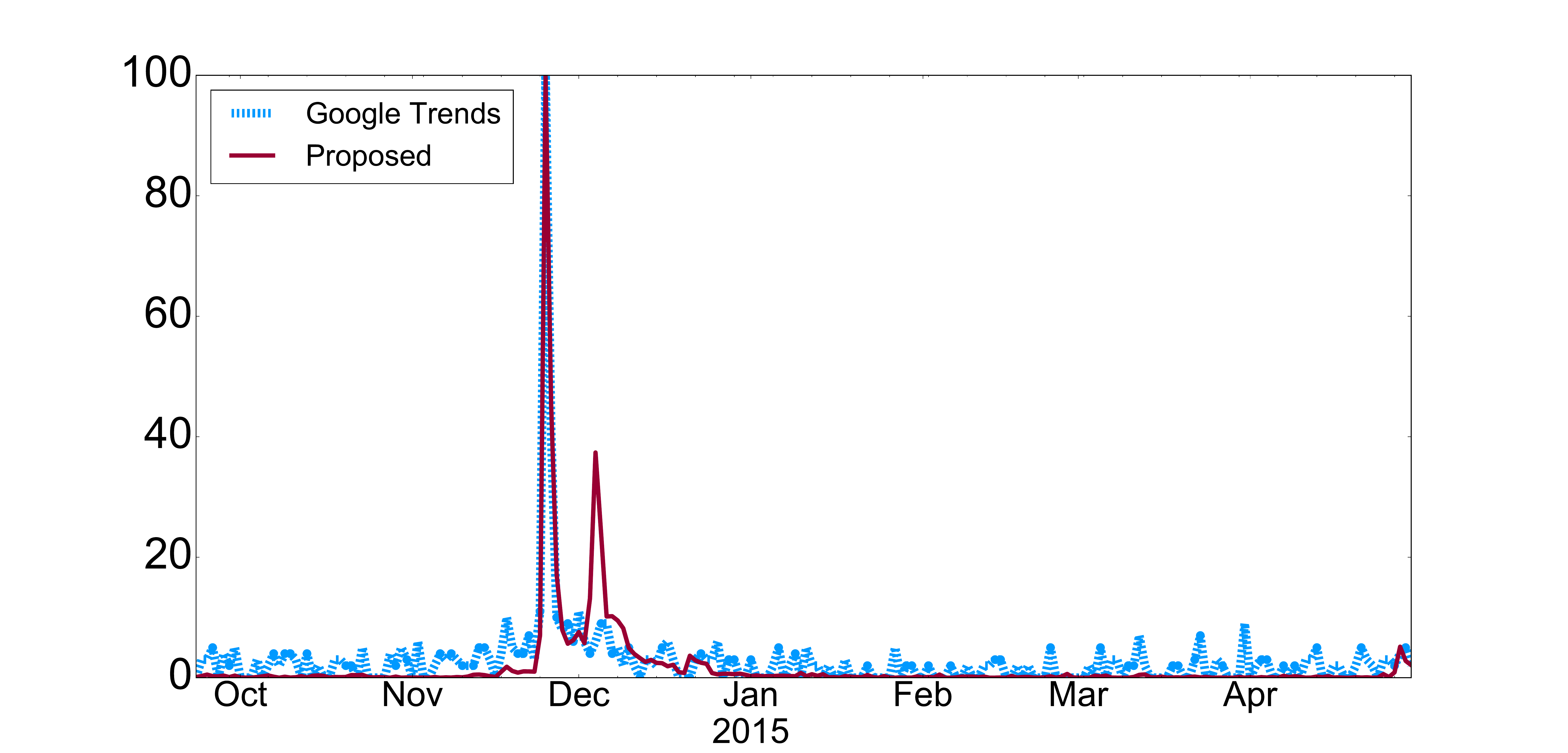} &
		\includegraphics[width=0.315\textwidth, trim={5cm 0cm 5cm 0cm}, clip] {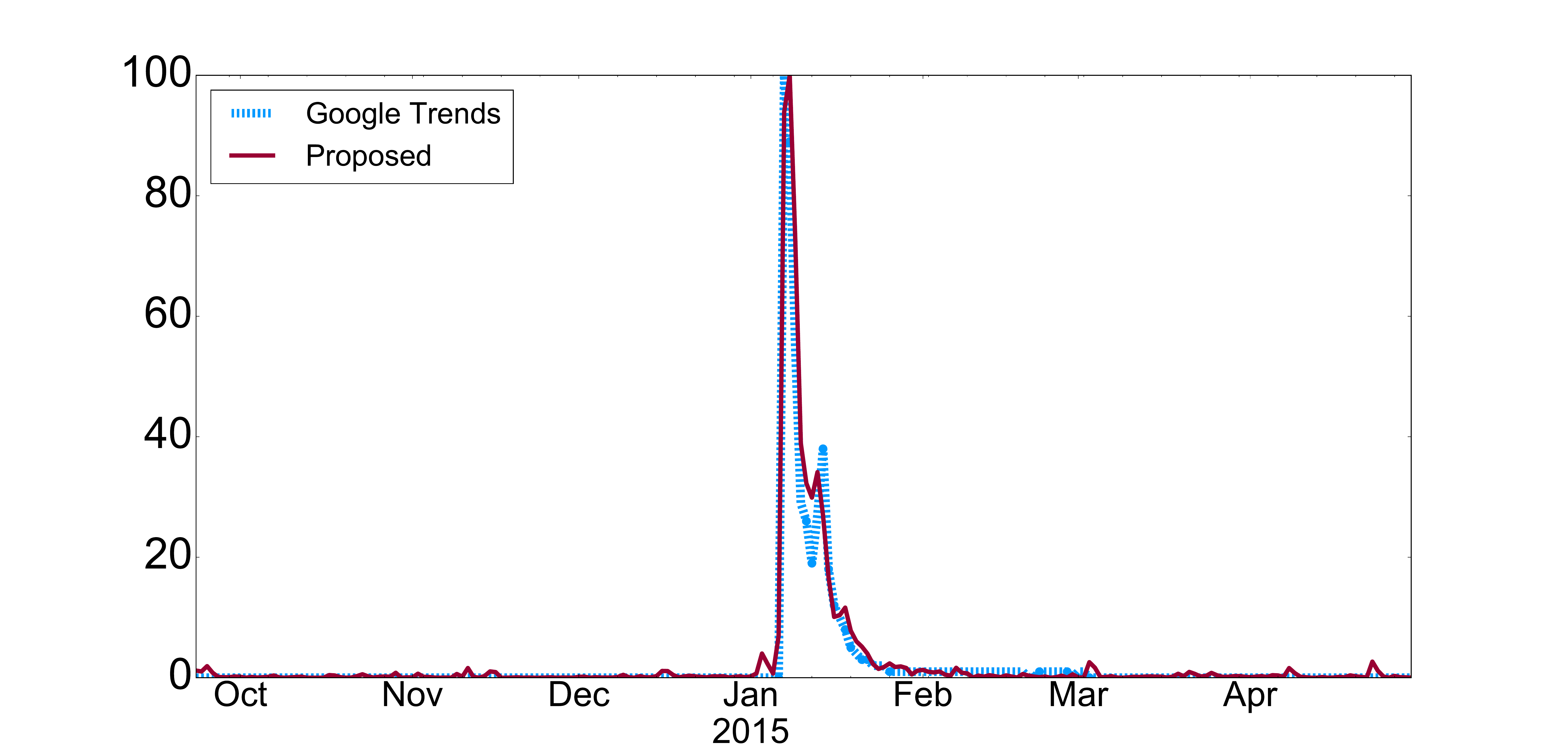} \\
		(a) Germanwings 9525 crash & (b) Ferguson unrest & (c) Charlie Hebdo attack
	\end{tabular}
	\caption{Graphs and activity timelines of the 3 events that triggered anomalies in the network dynamics. Top: Clusters of pages grouped after learning in the Hopfield network. Bottom: A normalized sum of all visits of the articles of each cluster over time (in red). The Google Trends curve for the keywords "Germanwings 9525 crash" (a), "Ferguson unrest" (b) and "Charlie Hebdo attack" (c) is displayed in blue.}
	\label{fig: dynamics}
\end{figure*}

We pick three core events among 172 detected and discuss them to show the details of our anomaly detection approach. Figure~\ref{fig: dynamics} shows the extracted clusters from the learned graph (top) and the overall timeline of the clusters' activity (bottom). We evaluate the accuracy of the anomaly detection using Google Trends history records.

\textit{Charlie Hebdo shooting.} 7 January 2015. This terrorist attack is an example of an unexpected event. The cluster emerged over a period of 72 hours, following the attack. All pages in the cluster are related to the core event. Strikingly, a look at the title of the pages is sufficient to get a precise summary of what the event is about. There is a sharp peak of activity on the first day of the attack, slowly decaying over the following week.

\textit{Germanwings flight 9525 crash.} 24 March 2015. This cluster not only involves pages describing the crash or providing more information about it, but also the pages of similar events that happened in the past. It includes, for example, a page enumerating airlane crashes and the page of a previous crash that happened in December 2014, the Indonesia AirAsia Flight 8501 crash. As a result, the time activity of the event is connected to the one of the Flight 8501 crash, that is why we can see an increase in visits in December. This is an example where our approach captures two relevant events and groups them together in one cluster allowing to detect a secondary anomaly that is only implicitly related to the main one.

\textit{Ferguson unrest. Second wave.} November 24, 2014 -- December 2, 2014. This is an example of an event that has the official beginning and end dates. A sharp increase in the activity at the beginning of protests highlights the main event. This moment triggers the core cluster emergence. We also see that the cluster becomes active once again at the end of the unrest allowing to record the two related anomalies in the visitor activity.


Finally, in Table \ref{table: collective_memories}, we summarize our exploration of the clusters of the learned graphs by providing a list of handpicked page titles inside each cluster that refer to previous events and related subjects. The connected events occurred outside of the 7-months period we analyze. This illustrates the associative property of our the anomalies we detect. Firstly, pages are grouped by events with the help of visitors activity. Secondly, the events themselves are connected together by this activity. An anomaly is characterized by a group of connected concepts and these groups are in turn connected together through concepts they share. This feature makes it easy to interpret the cause of the detected anomaly.

\noindent{\bf Google Trends Evaluation}. For each of the events presented before, we have compared the total number of visits in the clusters with Google Trends curves reflecting anomalous search activity of internet users. There is a striking correspondence between the detected anomalies and Google Trends, as we can see in Fig.~\ref{fig: SB_evolution} and~\ref{fig: dynamics}. In all 4 examples, the anomalous activity and Google Trends curves reach their maximum at the same time and have a very similar shape. The differences that appear during the months prior to the Super Bowl date are explained by the fact that our "Super Bowl" cluster contains articles about Football teams and other topics related to the Super Bowl. Due to the associative nature of the detected anomalies, this example has a higher accuracy of detection than Google Trends. 

As discussed previously, the periodic spikes in the visitor activity take place during the weekends when football matches are played. We observe the same phenomenon in the case of the Germanwings crash, where we observe a small peak of activity in December. This peak is the result of the prior activity on the pages related to another airplane crash that happened in December 2014. Again, this demonstrates the richness of the obtained clusters: they describe a group of events that caused an anomaly in visitor activity, as confirmed by Google Trends but represent more than a single keyword since the detected anomaly is a result of multiple real-world events.

\begin{table*}
	\caption{Examples of Wikipedia article titles contained in the clusters associated to the events presented.}
	\label{table: collective_memories}
	\begin{minipage}{\textwidth}
		\begin{center}
			\begin{tabular}{ lll }
				\toprule
				\textbf{Charlie Hebdo attack} & \textbf{Germanwings 9525 crash} & \textbf{Ferguson unrest} \\
				\midrule
				Porte de Vincennes hostage crisis & Inex-Adria Aviopromet Flight 1308 & Shooting of Tamir Rice \\
				Al-Qaeda & Pacific Southwest Airlines Flight 1771 & Shooting of Amadou Diallo \\
				Islamic terrorism & SilkAir Flight 185 & Sean Bell Shooting Incident \\ 
				Hezbollah & Suicide by pilot & Shooting of Oscar Grant \\
				2005 London bombings & Aviation safety & 1992 Los Angeles riots \\ 
				Anders Behring Breivik & Air France Flight 296 & O.J. Simpson murder case \\ 
				Jihadism & Air France Flight 447 & Shooting of Trayvon Martin \\ 
				2015 Baga massacre & Airbus & Attack on Reginald Denny \\
				\bottomrule
			\end{tabular}
		\end{center}
		\bigskip
	\end{minipage}
\end{table*}

\begin{figure*}
	\begin{tabular}{c}
		\includegraphics[width=\textwidth, trim={0cm 0cm 0cm 0cm}, clip]{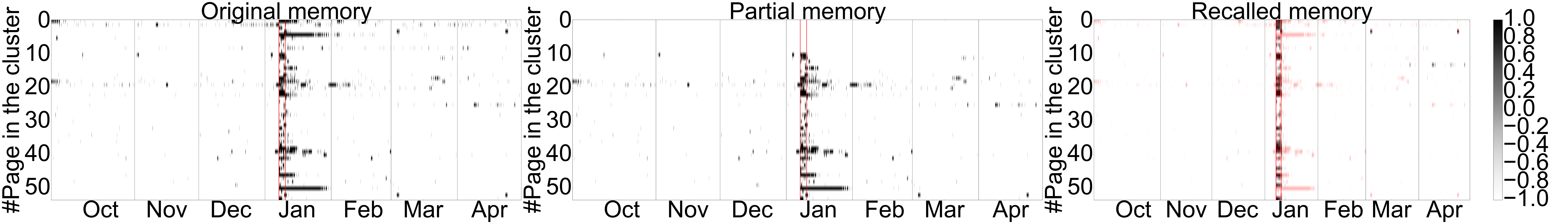}
	\end{tabular}
	\caption{Recall of an event from a partial pattern (Charlie Hebdo attack). The red vertical lines define the start of the event and its most active part, ending 72 hours from the start. Left: full activity over time of the pages in the cluster. Middle: pattern with 20\% of nodes set inactive (top lines). Right: the result of the recall using the Hopfield network model. In light red are shown the difference with the original pattern (the forgotten activity).}
	\label{fig: remind}
\end{figure*}

\noindent{\bf Recalling memories.}
In this last experiment, our goal is to test the hypothesis that the proposed method, as a memory, allows recalling or recovering events from partial information. We emulate recall processes using the Hopfield network approach, described in Section~\ref{Graph learning}. We show that the learned graph structure can recover a full event (cluster of pages and its activations) from an incomplete input.


We create incomplete patterns for the Hopfield network by selecting randomly a few pages contained in a chosen cluster. We built the input matrix $P_0$ setting to $(-1)$ (inactive state) all the time-series except for the few pages selected. We then apply iteratively Eq.~\eqref{eq: hopfield}.



We present only one example of the recall, to demonstrate the recall ability of the network but also to illustrate its complexity, on Fig.~\ref{fig: remind}.
From the cluster associated to the Charlie Hebdo Attack and we choose a subset of the list of articles, here $80\%$. We apply the learned graph for the month of January when the memory was detected. After the recall, we can remark two important facts. First, if we focus on the short time span when the event occurs (within the red vertical lines), most of the cluster is recovered. Second, the model forgets a part of the activity, plotted in light red, outside of the event bounds. This missing part is made of pages that are active outside of the time of the event, giving the evidence that they are not directly related (or weakly related) to the event.

The results of this experiments show that our method can be used to recover the signals related to the detected anomalies given a noisy or incomplete input. For instance, this feature can be helpful in cases when the data required for anomaly interpretation is destroyed by some intruders that are interested in hiding traces after an attack on the network.

\section{Conclusions and future work}\label{Conclusions}
In this paper, we presented a new unsupervised graph algorithm that allows detecting anomalies in large datasets with dynamic activity records. We showed that this approach allows analyzing the root causes of detected anomalies. Beyond the presentation of the method, we also provided new insights on collective behavior of Wikipedia users.

This approach is promising. We noted experimentally a high robustness of the method to the tuning of the parameters. We also plan to go deeper into the analysis of the recalling process.

This work opens new avenues for dynamic graph-structured data analysis. For example, the proposed approach could be used in a framework for automated event detection, monitoring, and recording. However, to cover these use cases, the method should be able to process streams of data as they are recorded. An online version of the method is not straightforward and will be the object of future work.

The artificial memory model used in the method suggests an interesting connection between the detection of anomalies and the functioning of the memory. This insight is more apparent in the experiments related to Wikipedia. Events and abnormal activity in the data trigger the recording process of the collective associative memory. The Hopfield network trained on Wikipedia articles transforms into a set of interconnected concepts or topics that have interested people at some point in the past. It resembles an artificial collective memory. This is fascinating and deserves further research in this direction related to social sciences.

\section{Tools, implementation, code and online visualizations} \label{reproducible}

All learned Wikipedia graphs (overall September-April, monthly activity, and localized events) are available online~\cite{WikiViz} to foster further exploration and analysis. For graph visualization we used the open source software package Gephi~\cite{bastian2009gephi} and layout ForceAtlas2~\cite{jacomy2014forceatlas2}. We used Apache Spark GraphX~\cite{xin2013graphx}, \cite{gonzalez2014graphx} for graph learning implementation and graph analysis. The presented results can be reproduced using the code, written in Scala and Python~\cite{Miz2017},~\cite{Miz2017enron}.
 Both datasets are available on Zenodo~\cite{benzi_kirell_2017_886484},~\cite{volodymyr_miz_2018_1342353}.

\begin{acks}
	
	We would like to thank Micha\"{e}l Defferrard and Andreas Loukas for fruitful discussions and useful suggestions. Also, we would like to thank anonymous reviewers for their constructive criticism and recommendations that helped improve this paper.
	
	The research leading to these results has received funding from the European Union's H2020 Framework Programme (H2020-MSCA-ITN-2014) under grant agreement n\textsuperscript{o}~642685 MacSeNet.
	
\end{acks}
\bibliographystyle{ACM-Reference-Format}
\bibliography{sample-bibliography}


\begin{thebibliography}{00}


\ifx \showCODEN    \undefined \def \showCODEN     #1{\unskip}     \fi
\ifx \showDOI      \undefined \def \showDOI       #1{#1}\fi
\ifx \showISBNx    \undefined \def \showISBNx     #1{\unskip}     \fi
\ifx \showISBNxiii \undefined \def \showISBNxiii  #1{\unskip}     \fi
\ifx \showISSN     \undefined \def \showISSN      #1{\unskip}     \fi
\ifx \showLCCN     \undefined \def \showLCCN      #1{\unskip}     \fi
\ifx \shownote     \undefined \def \shownote      #1{#1}          \fi
\ifx \showarticletitle \undefined \def \showarticletitle #1{#1}   \fi
\ifx \showURL      \undefined \def \showURL       {\relax}        \fi
\providecommand\bibfield[2]{#2}
\providecommand\bibinfo[2]{#2}
\providecommand\natexlab[1]{#1}
\providecommand\showeprint[2][]{arXiv:#2}

\bibitem[\protect\citeauthoryear{Aggarwal and Subbian}{Aggarwal and
  Subbian}{2014}]%
        {aggarwal2014evolutionary}
\bibfield{author}{\bibinfo{person}{Charu Aggarwal} {and}
  \bibinfo{person}{Karthik Subbian}.} \bibinfo{year}{2014}\natexlab{}.
\newblock \showarticletitle{Evolutionary network analysis: A survey}.
\newblock \bibinfo{journal}{{\em ACM Computing Surveys (CSUR)\/}}
  \bibinfo{volume}{47}, \bibinfo{number}{1} (\bibinfo{year}{2014}),
  \bibinfo{pages}{10}.
\newblock


\bibitem[\protect\citeauthoryear{Akoglu, Tong, and Koutra}{Akoglu
  et~al\mbox{.}}{2015}]%
        {akoglu2015graph}
\bibfield{author}{\bibinfo{person}{Leman Akoglu}, \bibinfo{person}{Hanghang
  Tong}, {and} \bibinfo{person}{Danai Koutra}.}
  \bibinfo{year}{2015}\natexlab{}.
\newblock \showarticletitle{Graph based anomaly detection and description: a
  survey}.
\newblock \bibinfo{journal}{{\em Data mining and knowledge discovery\/}}
  \bibinfo{volume}{29}, \bibinfo{number}{3} (\bibinfo{year}{2015}),
  \bibinfo{pages}{626--688}.
\newblock


\bibitem[\protect\citeauthoryear{Atluri, Karpatne, and Kumar}{Atluri
  et~al\mbox{.}}{2018}]%
        {atluri2018spatio}
\bibfield{author}{\bibinfo{person}{Gowtham Atluri}, \bibinfo{person}{Anuj
  Karpatne}, {and} \bibinfo{person}{Vipin Kumar}.}
  \bibinfo{year}{2018}\natexlab{}.
\newblock \showarticletitle{Spatio-temporal data mining: A survey of problems
  and methods}.
\newblock \bibinfo{journal}{{\em ACM Computing Surveys (CSUR)\/}}
  \bibinfo{volume}{51}, \bibinfo{number}{4} (\bibinfo{year}{2018}),
  \bibinfo{pages}{83}.
\newblock


\bibitem[\protect\citeauthoryear{Bastian, Heymann, Jacomy,
  et~al\mbox{.}}{Bastian et~al\mbox{.}}{2009}]%
        {bastian2009gephi}
\bibfield{author}{\bibinfo{person}{Mathieu Bastian}, \bibinfo{person}{Sebastien
  Heymann}, \bibinfo{person}{Mathieu Jacomy}, {et~al\mbox{.}}}
  \bibinfo{year}{2009}\natexlab{}.
\newblock \showarticletitle{Gephi: an open source software for exploring and
  manipulating networks.}
\newblock \bibinfo{journal}{{\em Icwsm\/}}  \bibinfo{volume}{8}
  (\bibinfo{year}{2009}), \bibinfo{pages}{361--362}.
\newblock


\bibitem[\protect\citeauthoryear{Benzi, Ricaud, and Vandergheynst}{Benzi
  et~al\mbox{.}}{2016}]%
        {benzi2016principal}
\bibfield{author}{\bibinfo{person}{Kirell Benzi}, \bibinfo{person}{Benjamin
  Ricaud}, {and} \bibinfo{person}{Pierre Vandergheynst}.}
  \bibinfo{year}{2016}\natexlab{}.
\newblock \showarticletitle{Principal Patterns on Graphs: Discovering Coherent
  Structures in Datasets.}
\newblock \bibinfo{journal}{{\em IEEE Trans. Signal and Information Processing
  over Networks\/}} \bibinfo{volume}{2}, \bibinfo{number}{2}
  (\bibinfo{year}{2016}), \bibinfo{pages}{160--173}.
\newblock


\bibitem[\protect\citeauthoryear{Benzi}{Benzi}{2017}]%
        {benzi2017recommender}
\bibfield{author}{\bibinfo{person}{Kirell~Ma{\"e}l Benzi}.}
  \bibinfo{year}{2017}\natexlab{}.
\newblock \showarticletitle{From recommender systems to spatio-temporal
  dynamics with network science}.
\newblock \bibinfo{publisher}{EPFL}, Chapter~5, \bibinfo{pages}{97--98}.
\newblock


\bibitem[\protect\citeauthoryear{Blondel, Guillaume, Lambiotte, and
  Lefebvre}{Blondel et~al\mbox{.}}{2008}]%
        {blondel2008fast}
\bibfield{author}{\bibinfo{person}{Vincent~D Blondel},
  \bibinfo{person}{Jean-Loup Guillaume}, \bibinfo{person}{Renaud Lambiotte},
  {and} \bibinfo{person}{Etienne Lefebvre}.} \bibinfo{year}{2008}\natexlab{}.
\newblock \showarticletitle{Fast unfolding of communities in large networks}.
\newblock \bibinfo{journal}{{\em Journal of statistical mechanics: theory and
  experiment\/}} \bibinfo{volume}{2008}, \bibinfo{number}{10}
  (\bibinfo{year}{2008}), \bibinfo{pages}{P10008}.
\newblock


\bibitem[\protect\citeauthoryear{Chandola, Banerjee, and Kumar}{Chandola
  et~al\mbox{.}}{2009}]%
        {chandola2009anomaly}
\bibfield{author}{\bibinfo{person}{Varun Chandola}, \bibinfo{person}{Arindam
  Banerjee}, {and} \bibinfo{person}{Vipin Kumar}.}
  \bibinfo{year}{2009}\natexlab{}.
\newblock \showarticletitle{Anomaly detection: A survey}.
\newblock \bibinfo{journal}{{\em ACM computing surveys (CSUR)\/}}
  \bibinfo{volume}{41}, \bibinfo{number}{3} (\bibinfo{year}{2009}),
  \bibinfo{pages}{15}.
\newblock


\bibitem[\protect\citeauthoryear{Diesner, Frantz, and Carley}{Diesner
  et~al\mbox{.}}{2005}]%
        {diesner2005communication}
\bibfield{author}{\bibinfo{person}{Jana Diesner}, \bibinfo{person}{Terrill~L
  Frantz}, {and} \bibinfo{person}{Kathleen~M Carley}.}
  \bibinfo{year}{2005}\natexlab{}.
\newblock \showarticletitle{Communication networks from the Enron email corpus
  ``It's always about the people. Enron is no different''}.
\newblock \bibinfo{journal}{{\em Computational \& Mathematical Organization
  Theory\/}} \bibinfo{volume}{11}, \bibinfo{number}{3} (\bibinfo{year}{2005}),
  \bibinfo{pages}{201--228}.
\newblock


\bibitem[\protect\citeauthoryear{Faghmous, Uluyol, Styles, Le, Mithal, Boriah,
  and Kumar}{Faghmous et~al\mbox{.}}{2013}]%
        {faghmous2013multiple}
\bibfield{author}{\bibinfo{person}{James~H Faghmous}, \bibinfo{person}{Muhammed
  Uluyol}, \bibinfo{person}{Luke Styles}, \bibinfo{person}{Matthew Le},
  \bibinfo{person}{Varun Mithal}, \bibinfo{person}{Shyam Boriah}, {and}
  \bibinfo{person}{Vipin Kumar}.} \bibinfo{year}{2013}\natexlab{}.
\newblock \showarticletitle{Multiple Hypothesis Object Tracking For
  Unsupervised Self-Learning: An Ocean Eddy Tracking Application.}. In
  \bibinfo{booktitle}{{\em AAAI}}.
\newblock


\bibitem[\protect\citeauthoryear{Ferron}{Ferron}{2012}]%
        {ferron2012collective}
\bibfield{author}{\bibinfo{person}{Michela Ferron}.}
  \bibinfo{year}{2012}\natexlab{}.
\newblock {\em \bibinfo{title}{Collective memories in Wikipedia}}.
\newblock \bibinfo{thesistype}{Ph.D. Dissertation}. \bibinfo{school}{University
  of Trento}.
\newblock


\bibitem[\protect\citeauthoryear{Ferron and Massa}{Ferron and Massa}{2011}]%
        {ferron2011studying}
\bibfield{author}{\bibinfo{person}{Michela Ferron} {and} \bibinfo{person}{Paolo
  Massa}.} \bibinfo{year}{2011}\natexlab{}.
\newblock \showarticletitle{Studying collective memories in Wikipedia}.
\newblock \bibinfo{journal}{{\em Journal of Social Theory\/}}
  \bibinfo{volume}{3}, \bibinfo{number}{4} (\bibinfo{year}{2011}),
  \bibinfo{pages}{449--466}.
\newblock


\bibitem[\protect\citeauthoryear{Foundation}{Foundation}{2016}]%
        {wikimediaPageCounts}
\bibfield{author}{\bibinfo{person}{Wikimedia Foundation}.}
  \bibinfo{year}{2016}\natexlab{}.
\newblock \bibinfo{title}{Wikimedia Downloads: Database tables as sql.gz and
  content as XML files}.
\newblock   (\bibinfo{date}{May} \bibinfo{year}{2016}).
\newblock
\showURL{%
\url{https://dumps.wikimedia.org/other/}}
\newblock
\shownote{Accessed: 2016-May-16.}


\bibitem[\protect\citeauthoryear{Garc{\'\i}a-Gavilanes, Mollgaard, Tsvetkova,
  and Yasseri}{Garc{\'\i}a-Gavilanes et~al\mbox{.}}{2017}]%
        {garcia2017memory}
\bibfield{author}{\bibinfo{person}{Ruth Garc{\'\i}a-Gavilanes},
  \bibinfo{person}{Anders Mollgaard}, \bibinfo{person}{Milena Tsvetkova}, {and}
  \bibinfo{person}{Taha Yasseri}.} \bibinfo{year}{2017}\natexlab{}.
\newblock \showarticletitle{The memory remains: Understanding collective memory
  in the digital age}.
\newblock \bibinfo{journal}{{\em Science Advances\/}} \bibinfo{volume}{3},
  \bibinfo{number}{4} (\bibinfo{year}{2017}), \bibinfo{pages}{e1602368}.
\newblock


\bibitem[\protect\citeauthoryear{Gonzalez, Xin, Dave, Crankshaw, Franklin, and
  Stoica}{Gonzalez et~al\mbox{.}}{2014}]%
        {gonzalez2014graphx}
\bibfield{author}{\bibinfo{person}{Joseph~E Gonzalez},
  \bibinfo{person}{Reynold~S Xin}, \bibinfo{person}{Ankur Dave},
  \bibinfo{person}{Daniel Crankshaw}, \bibinfo{person}{Michael~J Franklin},
  {and} \bibinfo{person}{Ion Stoica}.} \bibinfo{year}{2014}\natexlab{}.
\newblock \showarticletitle{GraphX: Graph Processing in a Distributed Dataflow
  Framework.}. In \bibinfo{booktitle}{{\em OSDI}}, Vol.~\bibinfo{volume}{14}.
  \bibinfo{pages}{599--613}.
\newblock


\bibitem[\protect\citeauthoryear{Griffa, Ricaud, Benzi, Bresson, Daducci,
  Vandergheynst, Thiran, and Hagmann}{Griffa et~al\mbox{.}}{2017}]%
        {griffa2017transient}
\bibfield{author}{\bibinfo{person}{Alessandra Griffa},
  \bibinfo{person}{Benjamin Ricaud}, \bibinfo{person}{Kirell Benzi},
  \bibinfo{person}{Xavier Bresson}, \bibinfo{person}{Alessandro Daducci},
  \bibinfo{person}{Pierre Vandergheynst}, \bibinfo{person}{Jean-Philippe
  Thiran}, {and} \bibinfo{person}{Patric Hagmann}.}
  \bibinfo{year}{2017}\natexlab{}.
\newblock \showarticletitle{Transient networks of spatio-temporal connectivity
  map communication pathways in brain functional systems}.
\newblock \bibinfo{journal}{{\em NeuroImage\/}}  \bibinfo{volume}{155}
  (\bibinfo{year}{2017}), \bibinfo{pages}{490--502}.
\newblock


\bibitem[\protect\citeauthoryear{Hebb}{Hebb}{2005}]%
        {hebb2005organization}
\bibfield{author}{\bibinfo{person}{Donald~Olding Hebb}.}
  \bibinfo{year}{2005}\natexlab{}.
\newblock \bibinfo{booktitle}{{\em The organization of behavior: A
  neuropsychological theory}}.
\newblock \bibinfo{publisher}{Psychology Press}.
\newblock


\bibitem[\protect\citeauthoryear{Hopfield}{Hopfield}{1982}]%
        {hopfield1982neural}
\bibfield{author}{\bibinfo{person}{John~J Hopfield}.}
  \bibinfo{year}{1982}\natexlab{}.
\newblock \showarticletitle{Neural networks and physical systems with emergent
  collective computational abilities}.
\newblock \bibinfo{journal}{{\em Proceedings of the national academy of
  sciences\/}} \bibinfo{volume}{79}, \bibinfo{number}{8}
  (\bibinfo{year}{1982}), \bibinfo{pages}{2554--2558}.
\newblock


\bibitem[\protect\citeauthoryear{Jacomy, Venturini, Heymann, and
  Bastian}{Jacomy et~al\mbox{.}}{2014}]%
        {jacomy2014forceatlas2}
\bibfield{author}{\bibinfo{person}{Mathieu Jacomy}, \bibinfo{person}{Tommaso
  Venturini}, \bibinfo{person}{Sebastien Heymann}, {and}
  \bibinfo{person}{Mathieu Bastian}.} \bibinfo{year}{2014}\natexlab{}.
\newblock \showarticletitle{ForceAtlas2, a continuous graph layout algorithm
  for handy network visualization designed for the Gephi software}.
\newblock \bibinfo{journal}{{\em PloS one\/}} \bibinfo{volume}{9},
  \bibinfo{number}{6} (\bibinfo{year}{2014}), \bibinfo{pages}{e98679}.
\newblock


\bibitem[\protect\citeauthoryear{Kanhabua, Nguyen, and Nieder{\'e}e}{Kanhabua
  et~al\mbox{.}}{2014}]%
        {kanhabua2014triggers}
\bibfield{author}{\bibinfo{person}{Nattiya Kanhabua}, \bibinfo{person}{Tu~Ngoc
  Nguyen}, {and} \bibinfo{person}{Claudia Nieder{\'e}e}.}
  \bibinfo{year}{2014}\natexlab{}.
\newblock \showarticletitle{What triggers human remembering of events? A
  large-scale analysis of catalysts for collective memory in Wikipedia}. In
  \bibinfo{booktitle}{{\em Digital Libraries (JCDL), 2014 IEEE/ACM Joint
  Conference on}}. IEEE, \bibinfo{pages}{341--350}.
\newblock


\bibitem[\protect\citeauthoryear{Kirell, Volodymyr, Benjamin, and
  Pierre}{Kirell et~al\mbox{.}}{2017}]%
        {benzi_kirell_2017_886484}
\bibfield{author}{\bibinfo{person}{Benzi Kirell}, \bibinfo{person}{Miz
  Volodymyr}, \bibinfo{person}{Ricaud Benjamin}, {and}
  \bibinfo{person}{Vandergheynst Pierre}.} \bibinfo{year}{2017}\natexlab{}.
\newblock \bibinfo{title}{Wikipedia time-series graph}.
\newblock   (\bibinfo{date}{Sept.} \bibinfo{year}{2017}).
\newblock
\showDOI{%
\url{https://doi.org/10.5281/zenodo.886484}}


\bibitem[\protect\citeauthoryear{Kratz and Nishino}{Kratz and Nishino}{2009}]%
        {kratz2009anomaly}
\bibfield{author}{\bibinfo{person}{Louis Kratz} {and} \bibinfo{person}{Ko
  Nishino}.} \bibinfo{year}{2009}\natexlab{}.
\newblock \showarticletitle{Anomaly detection in extremely crowded scenes using
  spatio-temporal motion pattern models}.
\newblock  (\bibinfo{year}{2009}).
\newblock


\bibitem[\protect\citeauthoryear{Lappas, Vieira, Gunopulos, and Tsotras}{Lappas
  et~al\mbox{.}}{2013}]%
        {lappas2013stem}
\bibfield{author}{\bibinfo{person}{Theodoros Lappas}, \bibinfo{person}{Marcos~R
  Vieira}, \bibinfo{person}{Dimitrios Gunopulos}, {and}
  \bibinfo{person}{Vassilis~J Tsotras}.} \bibinfo{year}{2013}\natexlab{}.
\newblock \showarticletitle{STEM: A spatio-temporal miner for bursty activity}.
  In \bibinfo{booktitle}{{\em Proceedings of the 2013 ACM SIGMOD International
  Conference on Management of Data}}. ACM, \bibinfo{pages}{1021--1024}.
\newblock


\bibitem[\protect\citeauthoryear{Li, Zang, Zhang, Li, and Wu}{Li
  et~al\mbox{.}}{2014}]%
        {li2014review}
\bibfield{author}{\bibinfo{person}{Miao Li}, \bibinfo{person}{Shuying Zang},
  \bibinfo{person}{Bing Zhang}, \bibinfo{person}{Shanshan Li}, {and}
  \bibinfo{person}{Changshan Wu}.} \bibinfo{year}{2014}\natexlab{}.
\newblock \showarticletitle{A review of remote sensing image classification
  techniques: The role of spatio-contextual information}.
\newblock \bibinfo{journal}{{\em European Journal of Remote Sensing\/}}
  \bibinfo{volume}{47}, \bibinfo{number}{1} (\bibinfo{year}{2014}),
  \bibinfo{pages}{389--411}.
\newblock


\bibitem[\protect\citeauthoryear{Lu, Kou, Zhao, and Chen}{Lu
  et~al\mbox{.}}{2007}]%
        {lu2007detecting}
\bibfield{author}{\bibinfo{person}{Chang-Tien Lu}, \bibinfo{person}{Yufeng
  Kou}, \bibinfo{person}{Jiang Zhao}, {and} \bibinfo{person}{Li Chen}.}
  \bibinfo{year}{2007}\natexlab{}.
\newblock \showarticletitle{Detecting and tracking regional outliers in
  meteorological data}.
\newblock \bibinfo{journal}{{\em Information Sciences\/}}
  \bibinfo{volume}{177}, \bibinfo{number}{7} (\bibinfo{year}{2007}),
  \bibinfo{pages}{1609--1632}.
\newblock


\bibitem[\protect\citeauthoryear{McEliece, Posner, Rodemich, and
  Venkatesh}{McEliece et~al\mbox{.}}{1987}]%
        {mceliece1987capacity}
\bibfield{author}{\bibinfo{person}{ROBERTJ McEliece}, \bibinfo{person}{EDWARDC
  Posner}, \bibinfo{person}{EUGENER Rodemich}, {and} \bibinfo{person}{SANTOSHS
  Venkatesh}.} \bibinfo{year}{1987}\natexlab{}.
\newblock \showarticletitle{The capacity of the Hopfield associative memory}.
\newblock \bibinfo{journal}{{\em IEEE transactions on Information Theory\/}}
  \bibinfo{volume}{33}, \bibinfo{number}{4} (\bibinfo{year}{1987}),
  \bibinfo{pages}{461--482}.
\newblock


\bibitem[\protect\citeauthoryear{Miz}{Miz}{2017a}]%
        {Miz2017}
\bibfield{author}{\bibinfo{person}{V. Miz}.} \bibinfo{year}{2017}\natexlab{a}.
\newblock \bibinfo{title}{Wikipedia graph mining. GraphX implementation}.
\newblock \bibinfo{howpublished}{\url{https://github.com/mizvol/WikiBrain}}.
  (\bibinfo{year}{2017}).
\newblock


\bibitem[\protect\citeauthoryear{Miz}{Miz}{2017b}]%
        {WikiViz}
\bibfield{author}{\bibinfo{person}{V. Miz}.} \bibinfo{year}{2017}\natexlab{b}.
\newblock \bibinfo{title}{Wikipedia graph mining. Visualization}.
\newblock   (\bibinfo{year}{2017}).
\newblock
\showURL{%
\url{https://goo.gl/Xgb7QG}}


\bibitem[\protect\citeauthoryear{Miz}{Miz}{2018}]%
        {Miz2017enron}
\bibfield{author}{\bibinfo{person}{V. Miz}.} \bibinfo{year}{2018}\natexlab{}.
\newblock \bibinfo{title}{Anomaly Detection in the Enron Email Graph. GraphX
  Implementation}.
\newblock
  \bibinfo{howpublished}{\url{https://github.com/mizvol/enron-email-network-analysis}}.
    (\bibinfo{year}{2018}).
\newblock


\bibitem[\protect\citeauthoryear{Miz, Ricaud, and Vandergheynst}{Miz
  et~al\mbox{.}}{2018}]%
        {volodymyr_miz_2018_1342353}
\bibfield{author}{\bibinfo{person}{Volodymyr Miz}, \bibinfo{person}{Benjamin
  Ricaud}, {and} \bibinfo{person}{Pierre Vandergheynst}.}
  \bibinfo{year}{2018}\natexlab{}.
\newblock \bibinfo{title}{Enron Email Time-Series Network}.
\newblock   (\bibinfo{date}{Aug.} \bibinfo{year}{2018}).
\newblock
\showDOI{%
\url{https://doi.org/10.5281/zenodo.1342353}}


\bibitem[\protect\citeauthoryear{Mongiovi, Bogdanov, Ranca, Papalexakis,
  Faloutsos, and Singh}{Mongiovi et~al\mbox{.}}{2013b}]%
        {mongiovi2013netspot}
\bibfield{author}{\bibinfo{person}{Misael Mongiovi}, \bibinfo{person}{Petko
  Bogdanov}, \bibinfo{person}{Razvan Ranca}, \bibinfo{person}{Evangelos~E
  Papalexakis}, \bibinfo{person}{Christos Faloutsos}, {and}
  \bibinfo{person}{Ambuj~K Singh}.} \bibinfo{year}{2013}\natexlab{b}.
\newblock \showarticletitle{Netspot: Spotting significant anomalous regions on
  dynamic networks}. In \bibinfo{booktitle}{{\em Proceedings of the 2013 SIAM
  International Conference on Data Mining}}. SIAM, \bibinfo{pages}{28--36}.
\newblock


\bibitem[\protect\citeauthoryear{Mongiovi, Bogdanov, and Singh}{Mongiovi
  et~al\mbox{.}}{2013a}]%
        {mongiovi2013mining}
\bibfield{author}{\bibinfo{person}{Misael Mongiovi}, \bibinfo{person}{Petko
  Bogdanov}, {and} \bibinfo{person}{Ambuj~K Singh}.}
  \bibinfo{year}{2013}\natexlab{a}.
\newblock \showarticletitle{Mining evolving network processes}. In
  \bibinfo{booktitle}{{\em Data Mining (ICDM), 2013 IEEE 13th International
  Conference on}}. IEEE, \bibinfo{pages}{537--546}.
\newblock


\bibitem[\protect\citeauthoryear{Ranshous, Shen, Koutra, Harenberg, Faloutsos,
  and Samatova}{Ranshous et~al\mbox{.}}{2015}]%
        {ranshous2015anomaly}
\bibfield{author}{\bibinfo{person}{Stephen Ranshous}, \bibinfo{person}{Shitian
  Shen}, \bibinfo{person}{Danai Koutra}, \bibinfo{person}{Steve Harenberg},
  \bibinfo{person}{Christos Faloutsos}, {and} \bibinfo{person}{Nagiza~F
  Samatova}.} \bibinfo{year}{2015}\natexlab{}.
\newblock \showarticletitle{Anomaly detection in dynamic networks: a survey}.
\newblock \bibinfo{journal}{{\em Wiley Interdisciplinary Reviews: Computational
  Statistics\/}} \bibinfo{volume}{7}, \bibinfo{number}{3}
  (\bibinfo{year}{2015}), \bibinfo{pages}{223--247}.
\newblock


\bibitem[\protect\citeauthoryear{Tinati, Luczak-Roesch, and Hall}{Tinati
  et~al\mbox{.}}{2016}]%
        {tinati2016finding}
\bibfield{author}{\bibinfo{person}{Ramine Tinati}, \bibinfo{person}{Markus
  Luczak-Roesch}, {and} \bibinfo{person}{Wendy Hall}.}
  \bibinfo{year}{2016}\natexlab{}.
\newblock \showarticletitle{Finding Structure in Wikipedia Edit Activity: An
  Information Cascade Approach}. In \bibinfo{booktitle}{{\em Proceedings of the
  25th International Conference Companion on World Wide Web}}. International
  World Wide Web Conferences Steering Committee, \bibinfo{pages}{1007--1012}.
\newblock


\bibitem[\protect\citeauthoryear{Wang, Tang, Park, and Priebe}{Wang
  et~al\mbox{.}}{2014}]%
        {wang2014locality}
\bibfield{author}{\bibinfo{person}{Heng Wang}, \bibinfo{person}{Minh Tang},
  \bibinfo{person}{Youngser Park}, {and} \bibinfo{person}{Carey~E Priebe}.}
  \bibinfo{year}{2014}\natexlab{}.
\newblock \showarticletitle{Locality statistics for anomaly detection in time
  series of graphs}.
\newblock \bibinfo{journal}{{\em IEEE Transactions on Signal Processing\/}}
  \bibinfo{volume}{62}, \bibinfo{number}{3} (\bibinfo{year}{2014}),
  \bibinfo{pages}{703--717}.
\newblock


\bibitem[\protect\citeauthoryear{Wu, Liu, and Chawla}{Wu et~al\mbox{.}}{2010}]%
        {wu2010spatio}
\bibfield{author}{\bibinfo{person}{Elizabeth Wu}, \bibinfo{person}{Wei Liu},
  {and} \bibinfo{person}{Sanjay Chawla}.} \bibinfo{year}{2010}\natexlab{}.
\newblock \showarticletitle{Spatio-temporal outlier detection in precipitation
  data}.
\newblock In \bibinfo{booktitle}{{\em Knowledge discovery from sensor data}}.
  \bibinfo{publisher}{Springer}, \bibinfo{pages}{115--133}.
\newblock


\bibitem[\protect\citeauthoryear{Xin, Gonzalez, Franklin, and Stoica}{Xin
  et~al\mbox{.}}{2013}]%
        {xin2013graphx}
\bibfield{author}{\bibinfo{person}{Reynold~S Xin}, \bibinfo{person}{Joseph~E
  Gonzalez}, \bibinfo{person}{Michael~J Franklin}, {and} \bibinfo{person}{Ion
  Stoica}.} \bibinfo{year}{2013}\natexlab{}.
\newblock \showarticletitle{Graphx: A resilient distributed graph system on
  spark}. In \bibinfo{booktitle}{{\em First International Workshop on Graph
  Data Management Experiences and Systems}}. ACM, \bibinfo{pages}{2}.
\newblock


\bibitem[\protect\citeauthoryear{Yu, Aggarwal, Ma, and Wang}{Yu
  et~al\mbox{.}}{2013}]%
        {yu2013anomalous}
\bibfield{author}{\bibinfo{person}{Weiren Yu}, \bibinfo{person}{Charu~C
  Aggarwal}, \bibinfo{person}{Shuai Ma}, {and} \bibinfo{person}{Haixun Wang}.}
  \bibinfo{year}{2013}\natexlab{}.
\newblock \showarticletitle{On anomalous hotspot discovery in graph streams}.
  In \bibinfo{booktitle}{{\em Data Mining (ICDM), 2013 IEEE 13th International
  Conference on}}. IEEE, \bibinfo{pages}{1271--1276}.
\newblock


\bibitem[\protect\citeauthoryear{Yucesoy and Barab{\'a}si}{Yucesoy and
  Barab{\'a}si}{2016}]%
        {yucesoy2016untangling}
\bibfield{author}{\bibinfo{person}{Burcu Yucesoy} {and}
  \bibinfo{person}{Albert-L{\'a}szl{\'o} Barab{\'a}si}.}
  \bibinfo{year}{2016}\natexlab{}.
\newblock \showarticletitle{Untangling performance from success}.
\newblock \bibinfo{journal}{{\em EPJ Data Science\/}} \bibinfo{volume}{5},
  \bibinfo{number}{1} (\bibinfo{year}{2016}), \bibinfo{pages}{17}.
\newblock


\end{thebibliography}

\end{document}